\documentclass[11pt]{article}
\usepackage[table]{xcolor}

\usepackage{fullpage, amsmath, amsthm, amsfonts, rotating}

\usepackage{geometry}

\usepackage{bm}
\usepackage{scalerel}

\onecolumn

\usepackage[T1]{fontenc}              
\usepackage[utf8]{inputenc}         
\usepackage{natbib}

\usepackage{amsthm,amsmath,amsfonts,amssymb} 
\usepackage{etoolbox}

\usepackage{graphicx}                 

\usepackage{listings}                 
\usepackage{tikz}
\usepackage{tikz-3dplot}
\usetikzlibrary{math} 
\usetikzlibrary{positioning}

\usepackage{algorithm2e}
\usepackage{mathtools}
\usepackage{enumerate}

\usepackage{epstopdf}
\usepackage{bbm}
\usepackage{placeins}
          
\usepackage{standalone}
\usepackage{url}
	
\usepackage{soul}

\theoremstyle{plain}

\theoremstyle{definition}

\theoremstyle{remark}

\usepackage{booktabs}
\usepackage{array}
\usepackage{multirow}
\usepackage{wrapfig}
\usepackage{float}
\usepackage{colortbl}
\usepackage{pdflscape}
\usepackage{tabu}
\usepackage{threeparttable}
\usepackage{threeparttablex}
\usepackage[normalem]{ulem}
\usepackage[utf8]{inputenc}
\usepackage{makecell}
\usepackage{verbatim}
\usepackage[symbol]{footmisc}

\title{Small Area Estimation of Fertility in Low- and Middle-Income Countries}
\date{\vspace{-5ex}}

\author{Yunhan Wu$^{1}$, Jon Wakefield$^{1,2}$\\
\vspace{0.2em}\\
$^1$Department of Biostatistics, University of Washington, Seattle, USA\\
$^2$Department of Statistics, University of Washington, Seattle, USA \phantom{te}}

\begin{document}

        \maketitle

\begin{abstract}
Accurate fertility estimates at fine spatial resolution are essential for localized public health planning, particularly in low- and middle-income countries (LMICs). While national-level indicators such as age-specific fertility rates (ASFR) and total fertility rate (TFR) are often reported through official statistics, they lack the spatial granularity needed to guide targeted interventions. To address this, we develop a framework for subnational fertility estimation using small-area estimation (SAE) techniques applied to birth history data from household surveys, in particular Demographic and Health Surveys (DHS). Disaggregation by geographic area, time period, and maternal age group leads to significant data sparsity, limiting the reliability of direct estimates at fine scales. To overcome this, we propose a suite of methods, including direct estimators, area-level and unit-level Bayesian hierarchical models, to produce accurate estimates across varying spatial resolutions. The model-based approaches incorporate spatiotemporal smoothing and integrate covariates such as maternal education, contraceptive use and urbanicity. Using data from the 2021 Madagascar DHS, we generate district-level ASFR and TFR estimates and evaluate model performance through cross-validation.

\end{abstract}

\section{Introduction}

In recent decades, global demographic shifts have led to a widespread decline in fertility \citep{ worldbank2010determinants, bongaarts2024fertility, finlay2018inequality}, with most countries reaching fertility rates at or below replacement levels \citep{bongaarts2020trends}. However, fertility remains persistently high in low- and middle-income countries (LMICs), averaging 4.7 births per woman in 2015--2020, nearly double the global average of 2.4 \citep{worldbank_fertility_2022}. High fertility, often correlated with economic and social factors, poses significant challenges, including risks to maternal and child health, reduced educational attainment, slowed economic growth, and increased pressure on environmental resources \citep{worldbank2010determinants}. Additionally, fertility rates influence disease burden projections, such as estimating the number of children at risk for conditions like HIV.

National fertility estimates, while useful for broad policy decisions, fail to capture local variations essential for effective intervention. Many reproductive health and family planning programs operate at the district (Admin-2) level, where precise fertility estimates are needed to guide resource allocation and service delivery \citep{mayala:etal:19}. Metrics such as the Age-Specific Fertility Rate (ASFR) and the Total Fertility Rate (TFR) provide critical insights into fertility levels and disparities. Reliable subnational fertility estimates enable targeted improvements in family planning and maternal health services \citep{saha2023small} and help policymakers identify inequities, optimize resource distribution, and design interventions tailored to community needs \citep{abate2024mapping}.

Localized fertility data are also essential for tracking progress toward Sustainable Development Goal (SDG) 3, Good Health and Well-Being, which prioritizes reducing maternal mortality and ensuring universal access to reproductive health services \citep{united2022sustainable}. SDG Target 3.7 further emphasizes the need for expanded access to contraception and reproductive healthcare, highlighting the importance of precise subnational fertility estimates for informed policy decisions and equitable service provision \citep{SDG4}. Moreover, fertility patterns also have significant implications for educational attainment, which is central to SDG 4, Quality Education. Specifically, SDG Target 4.1 aims to ensure that all girls and boys complete free, equitable, and quality primary and secondary education, an objective that can be influenced by fertility rates through factors such as adolescent pregnancies \citep{SDG4}.


Fertility estimation typically relies on birth history data, which is ideally recorded through Civil Registration and Vital Statistics (CRVS) systems \citep{adair2023global}. In high-income countries, probabilistic projections of total fertility rates have been conducted using long-term, reliable data sources \citep{vsevvcikova2018probabilistic}. However, these analyses are largely limited to regions with well-established CRVS systems. In contrast, most LMICs lack adequate CRVS coverage, making fertility estimation challenging and necessitating reliance on household surveys such as the Demographic and Health Surveys (DHS) and Multiple Indicator Cluster Surveys (MICS). 

Among these, DHS is one of the most widely used survey programs in LMICs, and collects extensive demographic, health, and social data across more than 90 countries and 300 surveys \citep{croft2018guide}. DHS and MICS surveys include georeferenced cluster locations, enabling subnational estimation, though historically geographical information is not available for MICS surveys. The methods we describe can be used for both surveys, but we focus on DHS. Survey data have been instrumental in estimating key demographic and health indicators, such as the under-five mortality rates (U5MR) and vaccination coverage  at the subnational level \citep{wakefield2019estimating,dong2021modeling}. In the context of fertility estimation, DHS surveys collect full birth histories from female respondents aged 15–49, including the timing of each child's birth. This information enables statistical models for fertility estimation to be developed.

While for generic indicators, DHS surveys provide reliable estimates at both the national and Admin-1 (first subnational administrative) levels, they often lack sufficient data to generate reliable estimates for smaller administrative areas (Admin-2) or finer geographic units. The challenge is further compounded by the nature of fertility data, which varies not only across space and time but also by the mother’s age group.
This additional decomposition, by mother's age, results in even sparser data, making it difficult to obtain reliable estimates at fine spatial and temporal resolutions.

To address these data challenges, small-area estimation (SAE) techniques provide a promising solution, by producing reliable estimates under data-sparse conditions. SAE approaches can be broadly categorized into design-based and model-based approaches \citep{rao:molina:15}. In general, design-based methods, such as direct estimation using survey weights, fully account for the complex survey design and provide design-consistent estimates. However, at fine spatial, temporal and demographic scales, these estimates become unreliable due to limited sample sizes, particularly in fertility estimation, where data sparsity is exacerbated across three dimensions—space, time, and maternal age group \citep{Ren2021}. Model-based approaches mitigate this issue by borrowing strength across areas and time, while also incorporating auxiliary covariates, leading to more precise estimates \citep{wakefield2020small}. However, for valid inference, model-based methods must account for DHS’s stratified two-stage cluster sampling; ignoring this design can bias estimates and lead to incorrect uncertainty quantification.


Despite the extensive use of SAE for health and demographic indicators in LMICs, no comprehensive workflow currently exists for applying SAE techniques to fertility estimation using household survey data. This study aims to bridge this gap by providing practical guidelines for SAE in fertility estimation, offering a structured framework with a statistically rigorous set of models tailored to different estimation goals and geographic levels.
Beyond improving subnational fertility estimation, our approach allows the incorporation of covariates in a predictive model framework, in particular allowing variables such as maternal education, contraceptive use and urbanicity to aid in prediction.


Maternal education plays a particularly influential role in shaping fertility patterns. Higher educational attainment is strongly associated with lower fertility, often through delayed marriage, increased labor force participation, and greater access to contraception and reproductive health knowledge \citep{martin1995women, jejeebhoy1995women, liu2024bayesian}. Regions with lower levels of female education, especially where secondary and higher education remains scarce, tend to have persistently high fertility rates \citep{pezzulo2021geographical}. In parallel, modern contraceptive use is also a key determinant of fertility outcomes, particularly in LMICs where unmet need for family planning remains substantial \citep{ibitoye2022fertility,sanchez2018adolescent}. Increased adoption of modern contraception is associated with fewer births and greater spacing between births \citep{bongaarts2017effect}.

These patterns suggest that using information on mother's education and contraceptive usage will help in SAE for fertility. We stress that in this paper we are interested in predicting fertility measures across time and space, and will use information on these two predictors within simple models; we do not attempt to elucidate the complex relationship between fertility and mother's education or contraceptive usage. 

Urban-rural differences further contribute to fertility disparities, as rural areas consistently exhibit higher fertility rates as compared to urban areas. While this gap is partially explained by differences in education, employment opportunities, and access to family planning, it persists even after adjusting for these factors, suggesting deeper structural and cultural influences \citep{ayele2015determinants, worldbank2010determinants}. Additionally, regional variations in fertility decline trends reflect broader demographic transitions in an increasingly urbanized world \citep{lerch2019regional}.
A key contribution of our framework is its ability to integrate  socioeconomic predictors into the SAE framework, giving the potential to obtain more precise estimates.


\begin{figure}[H]
    \centering
    \includegraphics[clip, trim=0.1cm 0.5cm 0.2cm 0.1cm, width=1\linewidth]{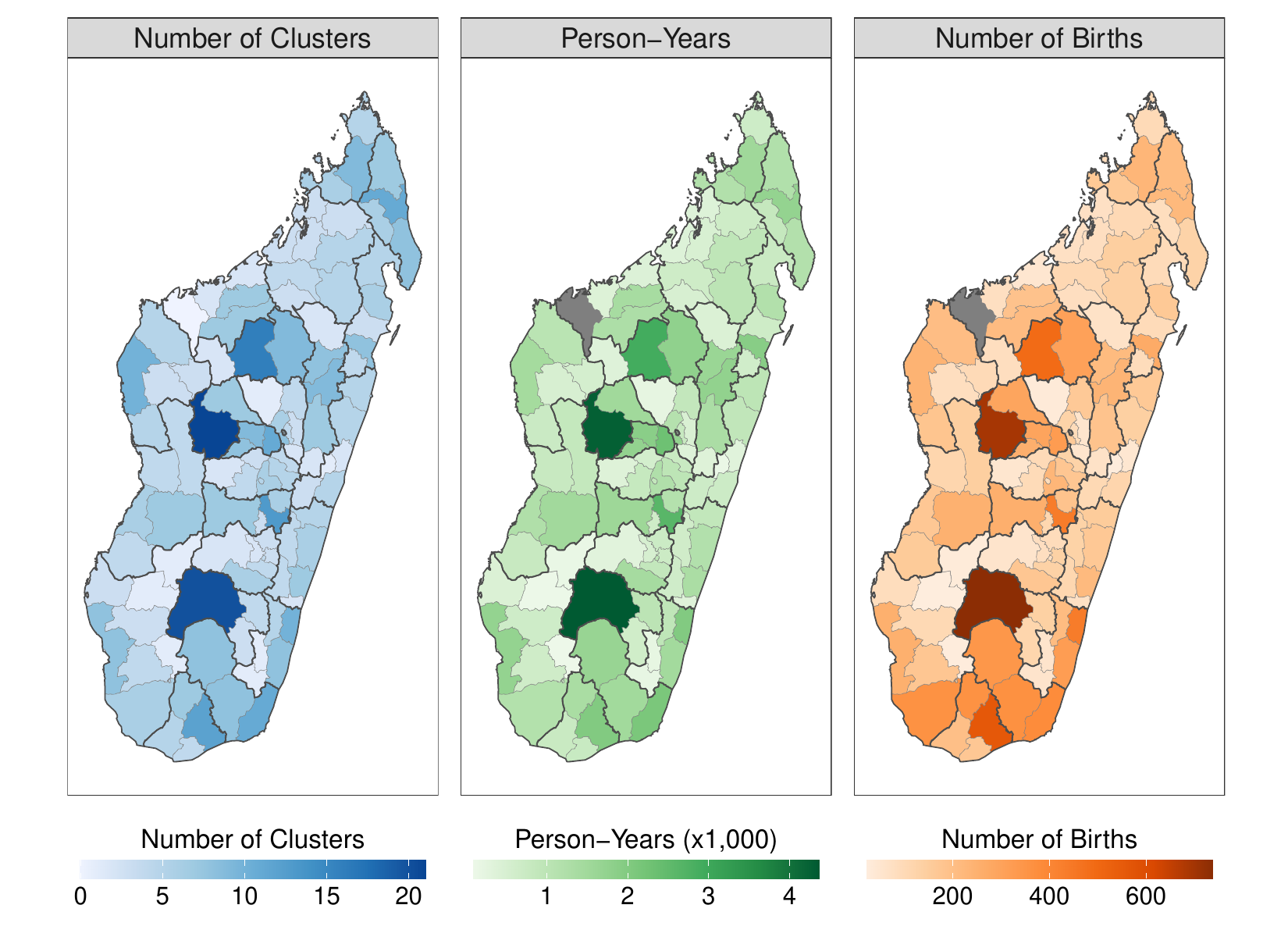}
\caption{Sample distribution across Admin-2 districts in Madagascar from the 2021 DHS. Maps show the number of sampled clusters, person-years contributed by women aged 15-49, and total births recorded between 2013 and 2021. One Admin-2 district, with no sampled clusters, appears as missing (shaded gray). Admin-1 boundaries are highlighted.}
\label{fig:sample-summary}
\end{figure}

To situate our methodological approach in a real-world context, we analyze fertility patterns over 2013--2021 in Madagascar using data from the 2021 DHS. The survey adopted a stratified, two-stage cluster sampling design. Stratification was defined by the intersection of urban-rural residence and Madagascar’s 23 administrative regions (Admin-1). Within each stratum, census enumeration areas, namely clusters, were selected in the first stage, using probability to size (PPS) sampling (with the size variable being the number of households). In the second stage of sampling, households were sampled within the clusters. In each cluster, 32 households were randomly selected and the total number of households that were planned for data collection was 21,024, with 5,248 located in urban areas and 15,776 in rural areas

We wish to estimate fertility rates in both the 23 Admin-1 regions and 119 Admin-2 districts. Women aged 15--49 contribute exposure time, measured in person-years, to the denominator of the fertility rate, with births during the reference period forming the numerator. The full 2021 DHS sample includes 647 clusters across the country. For this analysis, a small number of clusters were excluded due to missing geographic coordinates. The final analytic sample consists of 18,748 women, contributing 127,753 person-years of observation over the study period 2013--2021, during which 20,185 births were recorded. Figure~\ref{fig:sample-summary} presents the spatial distribution of the samples:~the number of sampled clusters, contributed person-years, and total births, all aggregated at the Admin-2 level. These patterns reveal considerable variation in data density across districts, necessitating our modeling frameworks that can borrow strength across geographic units to improve estimation in areas with sparse data.



The remainder of the paper is structured as follows. In Section \ref{sec:fert-method}, we present SAE methods, including direct estimation, area-level models, and unit-level models, to estimate ASFR and TFR based on data from a DHS survey. In Section \ref{sec:fert-res-data-application}, we demonstrate the application of our proposed models using the 2021 Madagascar DHS. Section \ref{sec:fert-cross-validation} evaluates the validity of the various modeling approaches through cross-validation. Finally, Section \ref{sec:fert-discussion} discusses our findings along with  future directions. Code to reproduce all results in this paper is available at {\tt https://github.com/wu-thomas/DHS-fertility}.

\section{Method}
\label{sec:fert-method}
We begin by defining the {\it age-specific fertility rate (ASFR)} and the {\it total fertility rate (TFR)}. We then employ a design-based direct estimation approach to illustrate the calculation of these metrics based on DHS survey data. Next, we introduce an area-level model to apply smoothing and enhance the precision of estimates by leveraging reliable direct estimates and their associated variances. Finally, we present a unit-level model, which provides, albeit with a greater number of assumptions, the most flexible framework, to achieve estimation at fine spatial resolution, such as Admin-2. This model also allows for the incorporation of individual-level covariates, such as maternal education.

\subsection{Definition of Fertility Metrics}

Both the ASFR and the TFR are time-dependent measures, meaning they are calculated with respect to a specific reference period. For simplicity, we do not explicitly state this reference period throughout, and we initially consider only national estimates -- space-time granularity will be introduced later.

The ASFR quantifies the frequency of childbearing among women within specific age groups. It is defined as the number of live births occurring in a given reference period per 1,000 women-years of exposure in the corresponding age group. In fertility estimation, age groups are typically divided into seven five-year intervals, denoted as $a \in  \mathcal{A}= \{
15-19, 20-24, 25-29, 30-34, 35-39, 40-44, 45-49\}$. 

Let \( Y_a \) denote the number of live births recorded among women in age group \( a \) during the reference period, and let \( E_a \) represent the corresponding total number of women-years of exposure. The ASFR for age group \( a \) is then computed as:
\begin{equation}
\text{ASFR}_{a} = \frac{Y_{a}}{E_{a}} \times 1000,\qquad a \in \mathcal{A}.
\end{equation}
The numerator \( Y_{a} \) is obtained by directly counting the number of births reported by women in age group \( a \) within the reference period. The denominator \( E_{a} \) is computed by summing the total women-years of exposure, accounting for transitions between age groups during the observation period. To ensure an accurate allocation of exposure time across age groups, birth dates for individual women are used to track age group progression within the reference window.

The TFR summarizes the fertility pattern across all reproductive ages by aggregating ASFR values over the seven age groups. Conceptually, the TFR represents the number of children who would be born per woman if she were to pass through the childbearing years bearing children according to the current schedule of age-specific fertility rates \citep{elkasabi2019calculating}. It is computed as:
\begin{equation}
\text{TFR} = 5 \times \sum_{a \in \mathcal{A}} \frac{\text{ASFR}_{a}}{1000}.
\end{equation}
The multiplication by 5 corresponds to the five-year width of each age group, such that the resulting estimate is the total number of births per woman over her lifetime under the current ASFR.

\subsection{Direct Estimation}
\label{sec:fert-direct-est-method}
DHS surveys collect complete birth histories from female respondents aged 15--49 years, and record the birth month and year of all of their children ever born. Additionally, the surveys include the respondent’s birth date, which allows us to track their transition across age groups over calendar time, as well as the geographic location associated with the survey cluster, which enables us to determine the administrative region within which the respondent belongs. Suppose we aim to estimate the ASFR for a given age group \( a \), in region \( i \), over a reference period \( t \) (e.g., three years preceding the survey) for $a \in \mathcal A$, $i=1,\dots,I$, $t=1,\dots,T$. Let \( j \in S_{i,t,a} \) represent the set of surveyed women who, at some point during reference period $t$, belonged to age group \( a \) in region \( i \). These women contribute to the calculation of \( \text{ASFR}_{i,t,a} \).

Since DHS employs a complex survey design involving stratification, clustering, and unequal probability sampling, each respondent $j$ is assigned a survey weight \( w_j \), which equals the inverse probability of selection (as well as a response rate adjustment). For woman $j$ contributing to the calculation of \( \text{ASFR}_{i,t,a} \), let \( y_{j} \) denote the number of births she had while in age group \( a \) during the reference period, and let \( E_{j} \) be the corresponding women-years of exposure.

The {\it weighted estimate} for \( \text{ASFR}_{i,t,a} \) is:
\begin{equation}
\widehat{\text{ASFR}}^{\mbox{w}}_{i,t,a} 
= \frac{\sum_{j \in S_{i,t,a}} w_{j} y_{j}}{\sum_{j \in S_{i,t,a}} w_{j}E_{j}} \times 1000.
\end{equation}
This is known as a direct estimate in the small area estimation (SAE) literature, since it is based on response data from the area $i$ and time period $t$ only.
The direct estimate for TFR for region \( i \) during reference period \( t \) is obtained by summing the direct ASFR estimates across age groups:
\begin{equation}
\widehat{\text{TFR}}^{\text{W}}_{i,t} = 5 \times \sum_{a \in \mathcal A} \frac{\widehat{\text{ASFR}}^{\text{W}}_{i, t,a}}{1000}.
\end{equation}


To quantify the uncertainty in ASFR and TFR estimates under the complex survey design, we employ replication-based variance estimation methods. The variance of these estimators does not have a closed-form expression and is typically estimated using either Taylor linearization or replication methods \citep{wolter2007introduction}. While Taylor linearization can approximate the variance for ratio estimators such as ASFR, it is not well-suited for composite indicators like TFR, whose variance cannot be easily decomposed analytically. Instead, replication methods such as the Jackknife offer a flexible, empirical approach that accounts for the complex structure of the estimator and provides valid uncertainty quantification \citep{pedersen2012child,elkasabi2019calculating}. To ensure consistency across ASFR and TFR, we adopt the Jackknife method for both. Estimation is implemented via the \textit{demogsurv} package in {\tt R} \citep{Eaton2020}.


\subsection{National ASFR Trends over Years}\label{sec:national}

\begin{figure}[H]
    \centering
    \includegraphics[clip, trim=0.1cm 0cm 0.2cm 0.1cm, width=1\linewidth]{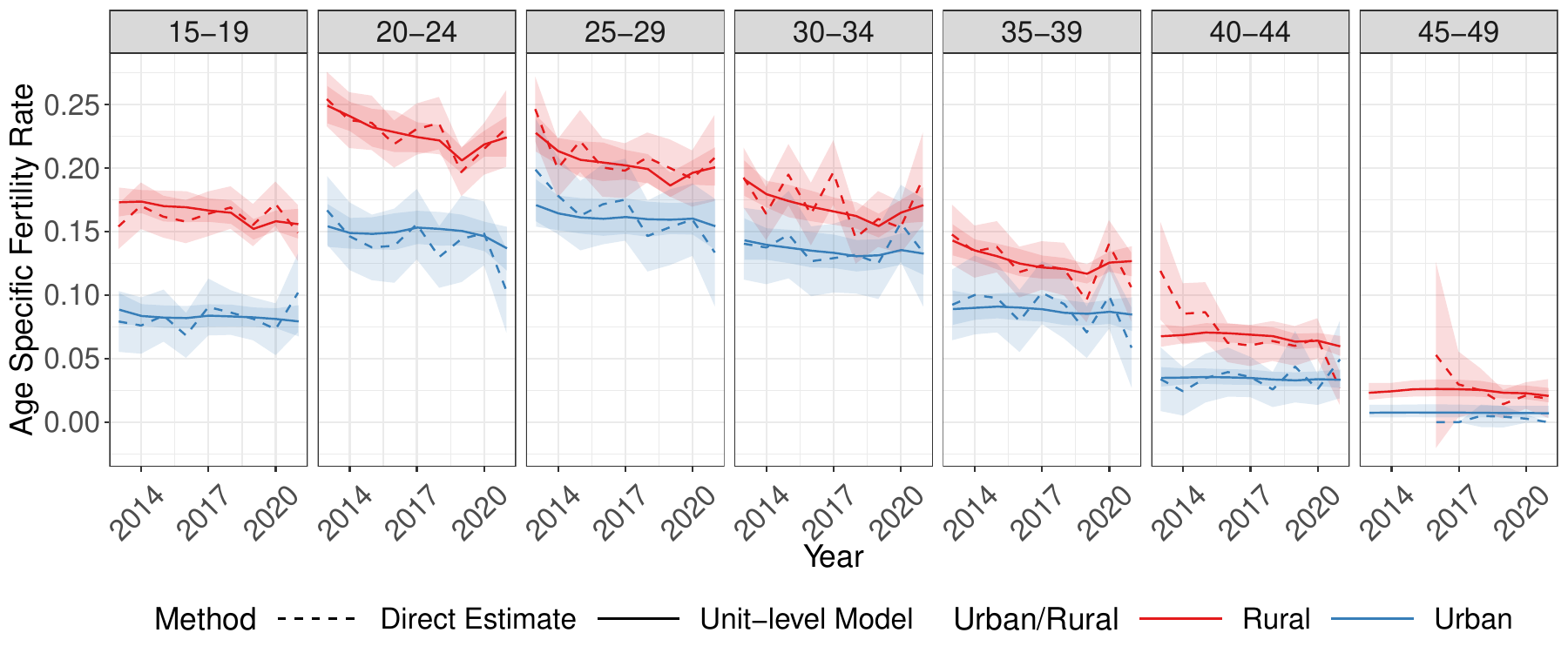}
\caption{National estimates of age-specific fertility rates (ASFR) over time (2013–2021) based on the 2021 Madagascar DHS, disaggregated by urban and rural populations.}
\label{fig:natl-yearly-ASFR-UR}
\end{figure}

To illustrate some aspects of the Madagascar data, we examine national trends. Figure 
\ref{fig:natl-yearly-ASFR-UR} presents national-level estimates of ASFR over time, disaggregated by urban and rural populations. These estimates are obtained using two methods:~the direct estimation approach just described in Section \ref{sec:fert-direct-est-method} and the unit-level small area model (with covariates) which we will describe  in Section \ref{sec:fert-unit-level-model}. The latter is implemented at the Admin-1 level and aggregated based on age-group-specific population.

Both urban and rural populations exhibit a declining trend in ASFR over time, particularly among the rural population. AS we will describe, the unit-level model produces smoother trends when compared to direct estimates, by reducing year-to-year fluctuations caused by sampling variability. Notice that for the oldest age group, we cannot calculate direct estimates for the early years, because of data sparsity. The observed temporal decline in ASFR is consistent with national level TFR estimates, which decrease from around 4.8 to 4.2 over the nine-year modeling period (2013--2021), aligning with estimates from World Population Prospects: The 2024 Revision \citep{UN_WPP_2024}.


Across all age groups, a persistent urban-rural fertility gap is evident, with rural areas consistently exhibiting higher ASFRs than urban areas. This disparity is most pronounced among younger reproductive ages (15--29), where rural fertility rates substantially exceed those in urban areas. These differences align with well-documented socioeconomic influences on fertility behavior, particularly access to education. Notably, the timing of peak fertility differs between urban and rural settings: in rural areas, fertility peaks at ages 20--24, whereas in urban areas, the peak occurs later, at ages 25–-29. This delayed fertility schedule in urban regions is likely driven by higher educational attainment and prolonged schooling.

Direct estimation is design-consistent and relies on minimal assumptions, making it a reliable approach when sufficient data are available, such as at the national or Admin-1 level. However, because it uses only data from each area/reference period, the associated uncertainties can be exceedingly high when sample sizes are small. In extreme cases, point estimates of 0 (no births) may occur (in the 45--49 age group, for example, as we see in Figure 
\ref{fig:natl-yearly-ASFR-UR}) and other sparse data situations can lead to instability in variance calculations. To address these limitations and improve precision in data-sparse areas/time periods, smoothing methods incorporated in area-level and unit-level models can be advantageous. These methods are introduced in the following sections.

\subsection{Area-level Model}
\label{sec:fert-area-level-model}

To improve precision for direct estimates, we apply the {\it area-level model} on a transformation of the direct estimates. This is an extended version of the Fay-Herriot model \citep{fay:herriot:79} that borrows strength across neighboring age groups, regions and reference periods.

\subsubsection{Area-level Model for ASFR}

We first introduce the area-level model for ASFR. We begin by applying a log transformation to the direct estimates to account for their rate-based nature \citep{chandra2011small}. Let \( \widehat{\eta}_{i,t,a} \) denote the log of the direct ASFR estimate in area \( i \) over reference period \( t \) for age group \( a \). Since direct estimates at fine temporal resolutions (e.g.,~yearly) are unreliable at the subnational level, we aggregate data into multi-year periods (e.g., three-year intervals) to improve stability. Given our focus on recent years and our wish  to not rely on older data (e.g.,~beyond 10 years before the survey), the number of available time points is limited, making temporal smoothing unfeasible. 

To simplify notation, we momentarily drop the reference period index \( t \). The associated design-based variance \( V_{i,a} \) is obtained using the delta method:
\begin{equation*}
 V_{i,a}= \frac{\widehat{\text{Var}}\left(\widehat{\text{ASFR}}^{\text{W}}_{i,a}\right)}{\left(\widehat{\text{ASFR}}_{i,a}\right)^2}.
\end{equation*}
The area-level model assumes that the sampling distribution of the transformed ASFR estimator is a normal distribution centered around the true (log) ASFR, incorporating measurement uncertainty from the survey design:
\begin{equation*}
\widehat{\eta}_{i,a} \mid \eta_{i,a} \sim N(\eta_{i,a}, V_{i,a}).
\end{equation*}
The justification is that this is the asymptotic distribution of the weighted estimator \citep{breidt2017model}. The linear predictor has the hierarchical structure,
\begin{eqnarray}
\eta_{i,a}  &=& 
\alpha +  \underbrace{\mathbf{x}_{i}^\top \boldsymbol{\beta}}_{\substack{
\text{Area-level}\\
\text{Covariate}}} +
  \underbrace{\phi_{a} + \psi_{a} }_{\substack{
\text{Mother's Age Group}\\
\text{Main Effects}
}}+ \underbrace{u_i }_{\substack{
\text{Spatial Effect} 
}} +
\underbrace{\delta^{(1)}_{i,a} }_{\substack{
\text{Space-Age Group}\\
\text{Interaction} 
}} 
\label{eq:FH-ASFR-model-mean}
\end{eqnarray}

\noindent where $\alpha$ is the intercept and other model components are defined as follows:

\begin{itemize}
    \item \textbf{Area-level Covariates:}
    $\mathbf{x}_{i}$ represents the area-level covariates included in the model. Specifically, we use the proportion of women aged 15-49 who have attended lower secondary school or higher, and the prevalence of modern contraceptive use, both estimated from the Madagascar 2021 DHS. To obtain stable estimates at both Admin-1 and Admin-2 levels, we apply a Fay-Herriot model \citep{fay:herriot:79} to direct survey estimates, incorporating a spatial random effect to improve precision in areas with sparse data (see Section~\ref{sec:covariate-model} in the supplemental material for details). The vector $\boldsymbol{\beta}$ contains the corresponding regression coefficients. We also consider a no-covariate version of the model, which excludes the $\mathbf{x}_{i}^\top \boldsymbol{\beta}$ term.

    
    \item \textbf{Mother’s Age Group Effects:} 

    \( \phi_{a} \) and \( \psi_{a} \) capture structured and unstructured effects associated with mother’s age group. The structured component \( \phi_{a} \) follows a first-order random walk (RW1) process to allow smoothing between adjacent age groups, while \( \psi_{a} \) represents an independent and identically distributed (IID) random effect.

    \item  \textbf{Spatial Effects for Areas:}

    \( u_i \) represents the spatial random effect for area \( i \). We adopt the BYM2 model \citep{riebler2016an}, a reparameterization of the Besag-York-Mollié (BYM) model \citep{besag1991bayesian}, which decomposes \( u_i \) into an IID component \( e_i \) and a spatially structured component \( S_i \):

    \begin{equation*}
    u_i = \sigma (\sqrt{1-\phi} e_i + \sqrt{\phi} S_i),
    \end{equation*}
where \( S_i \) follows a scaled intrinsic conditional autoregressive (ICAR) prior \citep{riebler2016an}. We use penalized complexity (PC) priors \citep{simpson2017penalising} for the hyperparameters \( \sigma \) (total standard deviation) and \( \phi \) (the proportion of the variation that is spatial).

    \item  \textbf{Space-Age Group Interaction:}

    The interaction term \( \delta^{(1)}_{i,a} \) accounts for deviations from the main effects of mother's age group and spatial components and is modeled as a type-IV space-time interaction \citep{knorr2000bayesian}.

\end{itemize}

\subsubsection{Area-level Model for TFR}

As TFR is an aggregate measure of ASFR, and does not include a mother’s age group component. In addition, we can obtain more reliable direct estimates at the yearly temporal resolution, at the Admin-1 level. This allows us to implement an area-level model that incorporates both spatial and temporal smoothing.

We use data from up to 9 years prior to the survey. Let \( \theta_{i,t} \) denote the log-transformed TFR for area \( i \) in year \( t \), with the direct estimate modeled as:
\begin{equation*}
\widehat{\theta}_{i,t} \mid \theta_{i,t} \sim N(\theta_{i,t}, V_{i,t}),
\end{equation*}
where \( V_{i,t} \) represents the design-based variance for \( \widehat{\theta}_{i,t} \).

The linear predictor is then modeled in a similar fashion, incorporating spatial and temporal dependencies to improve estimation precision.
\begin{eqnarray}
\theta_{i,t} &=& 
\alpha + \underbrace{I( t=t_s-6) \times \zeta -I( t=t_s-5) \times \zeta }_{\substack{
\text{Cutoff Bias Adjustment}
}} +
\underbrace{\mathbf{x}_{i}^\top \boldsymbol{\beta}}_{\substack{
\text{Area-level}\\
\text{Covariate}}} 
\nonumber \\
&+& \underbrace{\tau_t + \gamma_t}_{\substack{
\text{Temporal}\\
\text{Main Effects}
}}
+ \underbrace{u_i }_{\substack{
\text{Spatial Effect} 
}} +
\underbrace{\delta^{(2)}_{i,t} }_{\substack{
\text{Space-Time}\\
\text{Interaction} 
}} 
\label{eq:FH-TFR-model-mean}
\end{eqnarray}
where $\alpha$ is the intercept, $\mathbf{x}_{i}$ represents the area-level covariates, $\tau_t$ and $\gamma_t$ are the structured (RW2) and unstructured (IID) temporal trend in calendar time and $u_i$ are the spatial effect for area $i$. We use type-IV interactions between space and time $\delta^{(2)}_{i,t}$.

We introduce an adjustment term to address the cutoff year bias \citep{schoumaker2014quality,Eaton2020}, which arises from misrecorded birth age of the children. For children born within the five years before the survey, additional questions are included in the interview so that respondents and interviewers may record birth dates closer to the beginning of this five-year window, to avoid these extra questions. Consequently, we may observe an underestimation of fertility five years before the survey and an overestimation six years before the survey. Letting $t_s$ be the survey year, the shift parameter $\zeta$ relocates the misreported births to overcome the cutoff bias. However, with a single survey alone, the model can struggle to distinguish a change in the fertility rate (which is true signal) from the cutoff bias, so we assist the process by imposing the strong prior, $N(0.05, 0.1)$, on $\zeta$. 

Posterior inference for area-level models (and the unit-level models to be introduced later) is conducted using the Integrated Nested Laplace Approximation (INLA) method, implemented in the R package \texttt{INLA} \citep{rue2009approximate}.

The area-level model ensures design consistency by incorporating direct estimates while improving precision through smoothing. However, its key limitation is its reliance on design-based variance estimates associated with direct estimates, which becomes unstable at finer spatial and temporal resolutions. In particular, estimating ASFR for the 45--49 age group is challenging due to frequent zero birth counts in small samples, which results in variance estimation being unfeasible. Moreover, data sparsity prevents fully modeling the interactions between space, time, and mother’s age group. The model introduced next addresses these limitations by leveraging finer-scale data and a more flexible hierarchical structure.

\subsection{Model-based Unit-level Inference}
\label{sec:fert-unit-level-model}
Small area estimation (SAE) formulations that directly model individual outcomes are known as {\it unit-level models}. Since the ASFR represents the rate of births per mother-year, we model it as a count outcome with an exposure adjustment. In our approach, we aggregate individuals within each cluster who share the same characteristics.

Let \( Y_{c,t,a} \) denote the number of births from mothers in cluster \( c \), during year \( t \) (up to 9 years before the survey) and in age group \( a \in \mathcal A \). The total mother-years contributed, \( n_{c,t,a} \), serves as an offset.

Poisson models are a natural choice for count data with a rare outcome; however, to account for overdispersion (excess variation attributed to cluster sampling), we adopt a \textit{negative binomial model} with log link, which allows the variance to exceed the mean:

\begin{equation}\label{eq:fert-nb}
Y_{c,t,a} ~|~  \mu_{c,t,a},d \sim \text{NegBin}(\mu_{c,t,a}, d),
\end{equation}
with  \( \mu_{c,t,a} \) being the expected number of births for mothers in cluster \( c \), time \( t \), and in age group \( a \), given the exposure \( n_{c,t,a} \), with
\begin{equation*}
\mu_{c,t,a}  = n_{c,t,a} \times \exp(\eta_{c,t,a}).
\label{eq:fert-nb-link}
\end{equation*}
The overdispersion parameter \( d \) captures the excess variation beyond a Poisson process with marginal variance,
\begin{equation*}
\text{var}(Y_{c,t,a} ~|~  \mu_{c,t,a},d) = \mu_{c,t,a} \left( 1 + \frac{\mu_{c,t,a}}{d} \right),
\end{equation*}
so that large \( d \) corresponds to minimal overdispersion. 
The linear predictor is
\begin{eqnarray}
\eta_{c,t,a} &=& 
\alpha + \underbrace{I( t=t_s-6) \times \zeta -I( t=t_s-5) \times \zeta }_{\substack{
\text{Cutoff Bias Adjustment}
}} +
\underbrace{\mathbf{x}_{i}^\top \boldsymbol{\beta}}_{\substack{
\text{Area-level}\\
\text{Covariate}}} 
\nonumber \\
&+& \underbrace{\tau_t + \gamma_t}_{\substack{
\text{Temporal}\\
\text{Main Effects}
}}+
  \underbrace{\phi_{a} + \psi_{a} +}_{\substack{
\text{Mother's Age Group}\\
\text{Main Effects}
}} 
+\underbrace{u_{i[\mbox{\boldmath $s_c$}]} }_{\substack{
\text{Spatial Main Effects} 
}} 
\nonumber \\
&+& \underbrace{\delta^{(1)}_{i,a} }_{\substack{
\text{Space - Age group}\\
\text{Interaction} 
}} +   \underbrace{\delta^{(2)}_{i,t} }_{\substack{
\text{Space - Time}\\
\text{Interaction} 
}} + \underbrace{\delta^{(3)}_{a,t} }_{\substack{
\text{Age group - Time}\\
\text{Interaction} 
}} 
,\label{eq:fert-unit-ASFR-mean}
\end{eqnarray}
with the model terms defined as in \eqref{eq:FH-ASFR-model-mean} and \eqref{eq:FH-TFR-model-mean}. Since all individuals within the same area, age group, and reference window share the same linear predictor, the ASFR is obtained by exponentiating this predictor: $\text{ASFR}_{i,t,a} = \exp(\eta_{i,t,a})$. The TFR is then computed as a weighted sum of ASFR values across all maternal age groups.

\subsubsection{Urban/Rural Stratification}

The unit-level model defined in (\ref{eq:fert-nb}) does not incorporate survey design weights and, therefore, does not fully account for the complex sampling design. Ignoring urban/rural stratification can introduce significant biases \citep{wu2024modelling,dong2021modeling}, when:~(1) there is systematic over- or under-sampling of urban clusters (with the former being a common feature of DHS surveys), and (2) there are fertility differences between urban and rural areas. These disparities are particularly relevant in the estimation of ASFR and TFR, as fertility rates tend to be systematically lower in urban areas compared to rural.

To mitigate the bias and explicitly capture urban-rural fertility differences, we fit separate negative binomial models for urban and rural populations. We denote the corresponding fertility rates as \( \text{ASFR}_{i,t,a}^{\text{U}} \) and \( \text{ASFR}_{i,t,a}^{\text{R}} \), following the modeling framework described above. Estimating these models separately allows for a more accurate representation of the distinct fertility patterns in urban and rural areas, and also enables the examination of urban-rural disparities. 

Since the models are estimated independently for urban and rural clusters, an additional {\it aggregation step} is required to obtain an overall ASFR. The combined estimate is:
\begin{equation}
\mu_{i,t,a} = r_{i,t,a} \times \mu_{i,t,a}^{\text{U}} + (1 - r_{i,t,a}) \times \mu_{i,t,a}^{\text{R}},
\end{equation}
where \( r_{i,t,a} \) represents the proportion of the urban female population in area $i$, during reference period $t$ and age group $a$. To estimate \( r_{i,t,a} \), we adopt a recent approach that employs classification models to reconstruct urban/rural partitions \citep{wu2024modelling}. 

\section{Subnational Fertility Estimation in Madagascar}
\label{sec:fert-res-data-application}

In this section, we analyze the most recent DHS survey for Madagascar, which occurred in 2021, to provide  subnational fertility estimates. Our analysis examines fertility across three key dimensions: maternal age group decomposition (via ASFR), time, and most importantly (for SAE), space. Additionally, we assess urban-rural fertility disparities and investigate the impact of incorporating maternal education and modern contraceptive usage as covariates. While we account for temporal fertility trends (with 9-year worth of data) using the retrospective nature of full birth history data, our primary focus is to obtain reliable estimates for the most recent period (3 years preceding the survey) rather than studying long-term fertility trends.


\begin{figure}[H]
    \centering
    \includegraphics[clip, trim=0.2cm 0cm 0.2cm 0cm, width=1\linewidth]{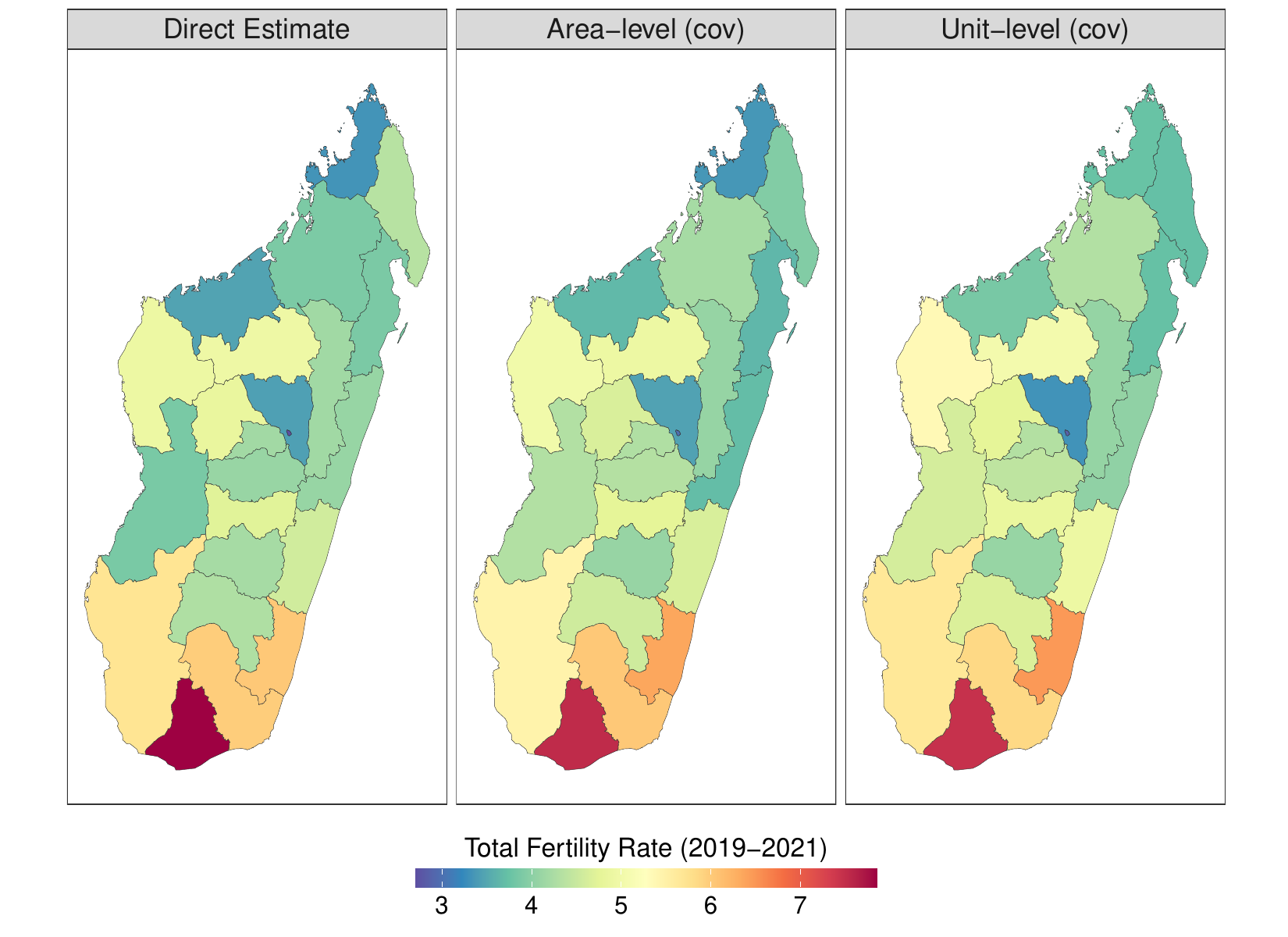}
    \caption{TFR estimates for Madagascar at the Admin-1 level for the most recent three-year period (2019--2021) across all three modeling approaches introduced in this SAE framework. The versions shown for area-level model and unit-level model incorporate area-level covariates.}
    \label{fig:adm1-TFR-all}
\end{figure}

\subsection{Subnational TFR Estimates}

In Section \ref{sec:national}, we examined national temporal trends.
We now shift our focus to finer spatial resolutions, extending our fertility analysis to Admin-1 (regional) and Admin-2 (district) levels. Madagascar consists of 23 Admin-1 regions, aligning with DHS survey stratification by singling out the capital city Antananarivo as a single region and merging the Fitovinany and Vatovavy regions. The 23 Admin-1 regions are further subdivided into 119 districts.

Figure \ref{fig:adm1-TFR-all} presents TFR estimates for the most recent three-year period (2019--2021) preceding the 2021 Madagascar DHS, comparing all modeling approaches considered:~the direct estimation method (Section \ref{sec:fert-direct-est-method}), the area-level model (Section \ref{sec:fert-area-level-model}), and the unit-level model (Section \ref{sec:fert-unit-level-model}). Each successive model introduces increased flexibility in capturing spatial relationships, but also additional modeling assumptions.

\begin{figure}[H]
    \centering
    \includegraphics[clip, trim=0.2cm 4.5cm 0.2cm 4.2cm, width=1\linewidth]{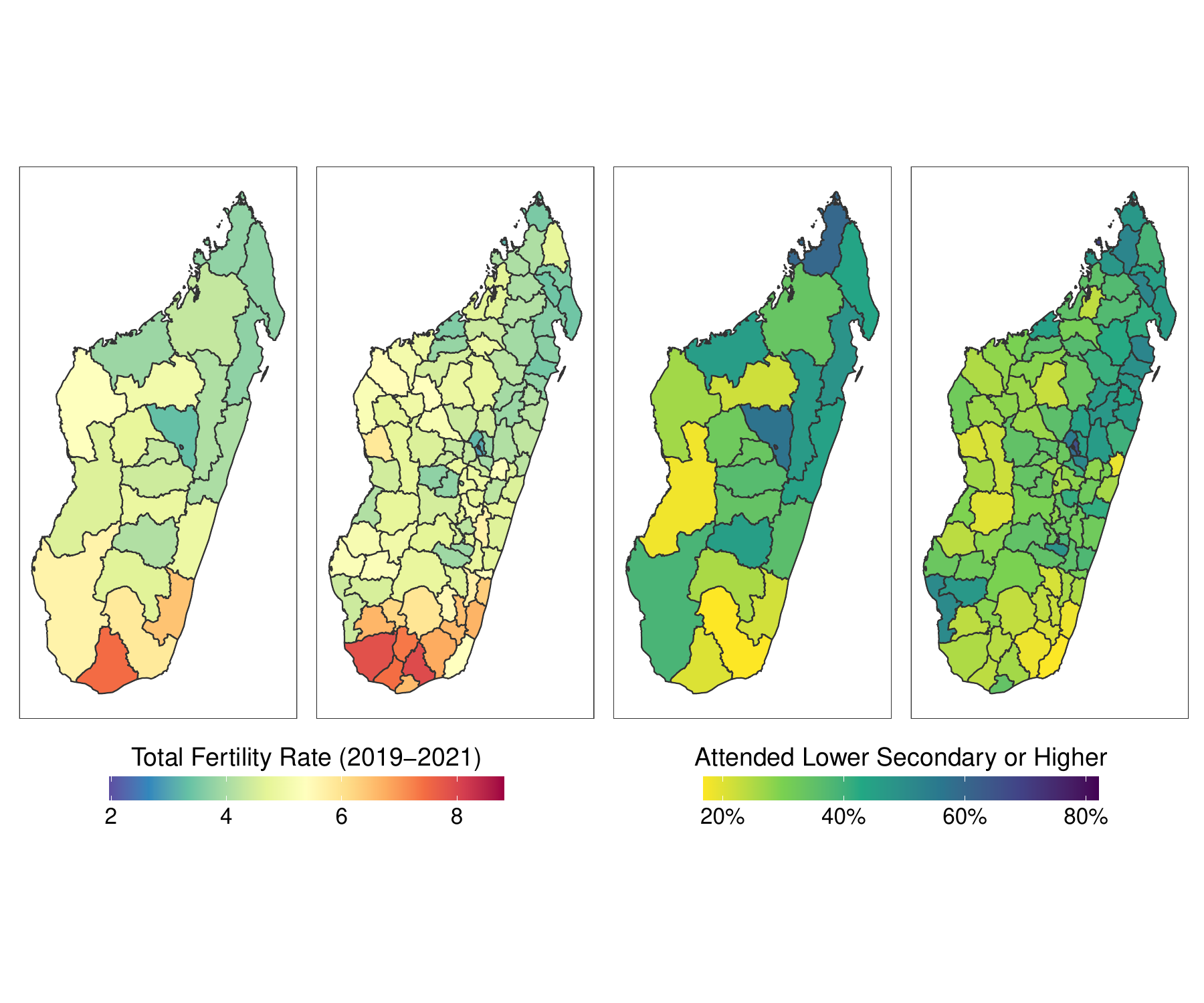}
    \caption{Subnational estimates of TFR (left panels) and maternal education attainment measured as percentage of attending secondary school (right panels) in Madagascar for the period 2019--2021. The maps display spatial variation at both Admin-1 and Admin-2 levels.}
    \label{fig:adm1-adm2-TFR-UYS}
\end{figure}




At the Admin-1 resolution, the available data are sufficient to yield reliable estimates from all four modeling methods, resulting in consistent TFR estimates across regions. The spatial patterns indicate regions with the highest fertility, notably Androy in the south, and regions with lower fertility, especially the capital region Antananarivo, situated in the center of Madagascar (though small in geographic area and so less visible on the map). Figure~\ref{fig:TFR-interval-adm1} in the supplemental material presents interval plots of point estimates and their associated uncertainties. Spatial smoothing effects are evident, particularly in the southernmost region, where the area-level and unit-level model estimates are drawn closer toward neighboring regions.

Figure \ref{fig:adm1-adm2-TFR-UYS} compares subnational estimates of TFR (from unit-level model with covariates) and maternal educational attainment (measured as percentage of women of age 15-49 attended secondary school or higher) at Admin-1 and Admin-2 levels. The Admin-2 level displays greater spatial variability compared to Admin-1, capturing more spatial heterogeneity in both fertility rates and educational attainment, emphasizing the importance of district-level analysis to uncover detailed spatial patterns. 

A clear pattern emerges: areas characterized by higher fertility consistently coincide with lower high school attendance rate, while regions with lower fertility typically exhibit higher levels of education. This inverse relationship between fertility and female education is visible at both administrative resolutions but is more nuanced at the district level. 

\begin{figure}[H]
    \centering
    \includegraphics[clip, trim=0.1cm 0.5cm 0.2cm 0.6cm, width=1\linewidth]{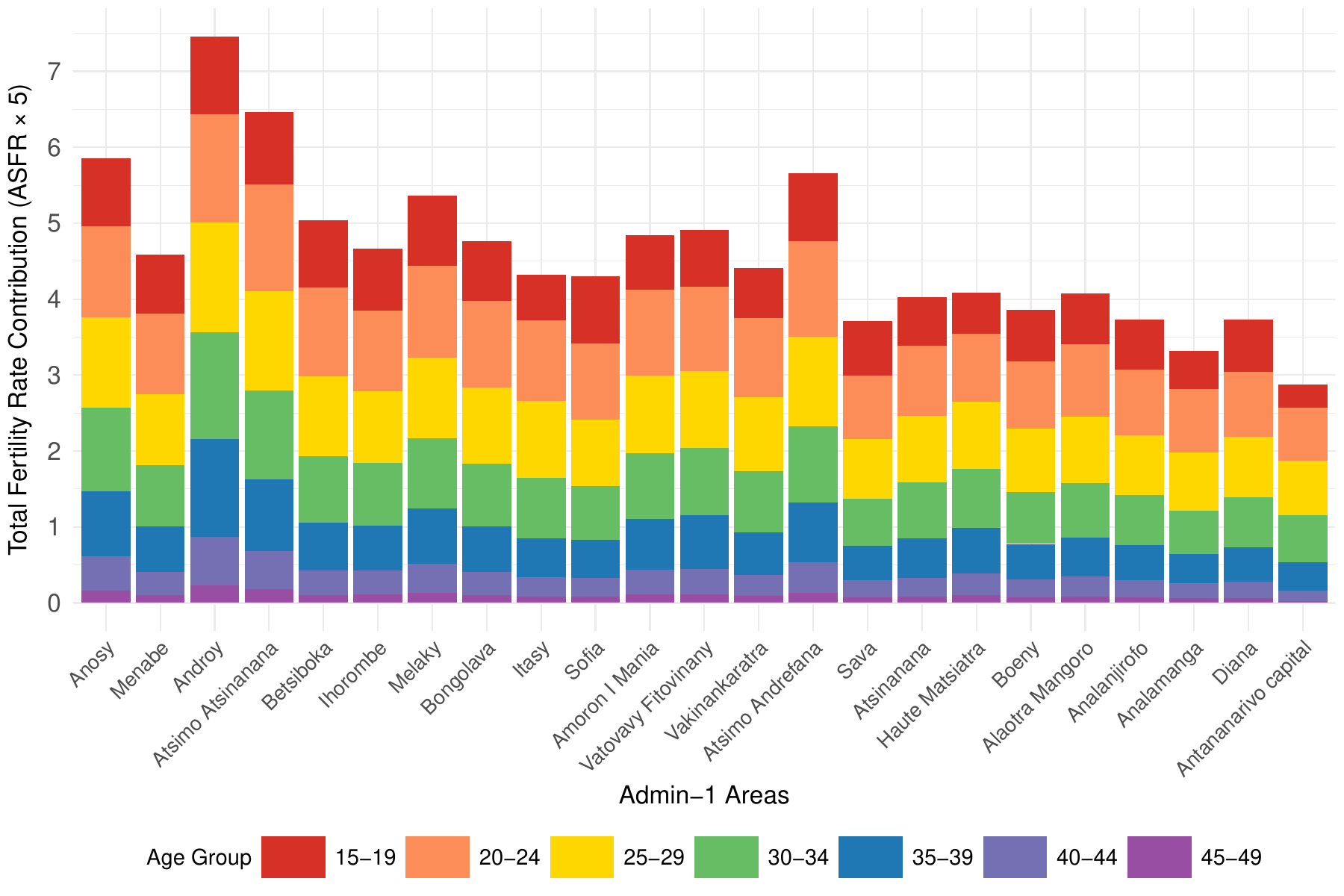}
    \caption{Decomposition of TFR by ASFR across Madagascar's Admin-1 regions (2019--2021). Regions are arranged from left to right according to increasing female high school attendance rate.}
    \label{fig:adm1-ASFR-bar}
\end{figure}
\subsection{Subnational ASFR Estimates}

Figure \ref{fig:adm1-ASFR-bar} illustrates regional variation in fertility across Madagascar's Admin-1 areas, ordered by increasing female educational attainment (measured as high school attendance rate). The estimates are from unit-level model with covariates. Each bar's height corresponds to the TFR of the region for the 3-year period preceding the survey (2019--2021), while the stacked segments indicate the contribution of each ASFR to the total fertility ($\text{height for each segment} = 5\times \text{ASFR}$).

Regions with lower female high school attendance, such as Anosy and Androy, exhibit higher fertility rates overall, while regions with higher education levels, such as the capital Antananarivo, display lower fertility. The age group specific decomposition further highlights regional differences in both fertility levels and the timing of childbearing, particularly in the contribution from the youngest age group (15–-19). These differences reflect subnational variation in fertility patterns across Madagascar.

\subsection{Urban-Rural Disparities in TFR}


\begin{figure}[ht]
    \centering
    \includegraphics[clip, trim=0.1cm 0.5cm 0.2cm 0.2cm, width=1\linewidth]{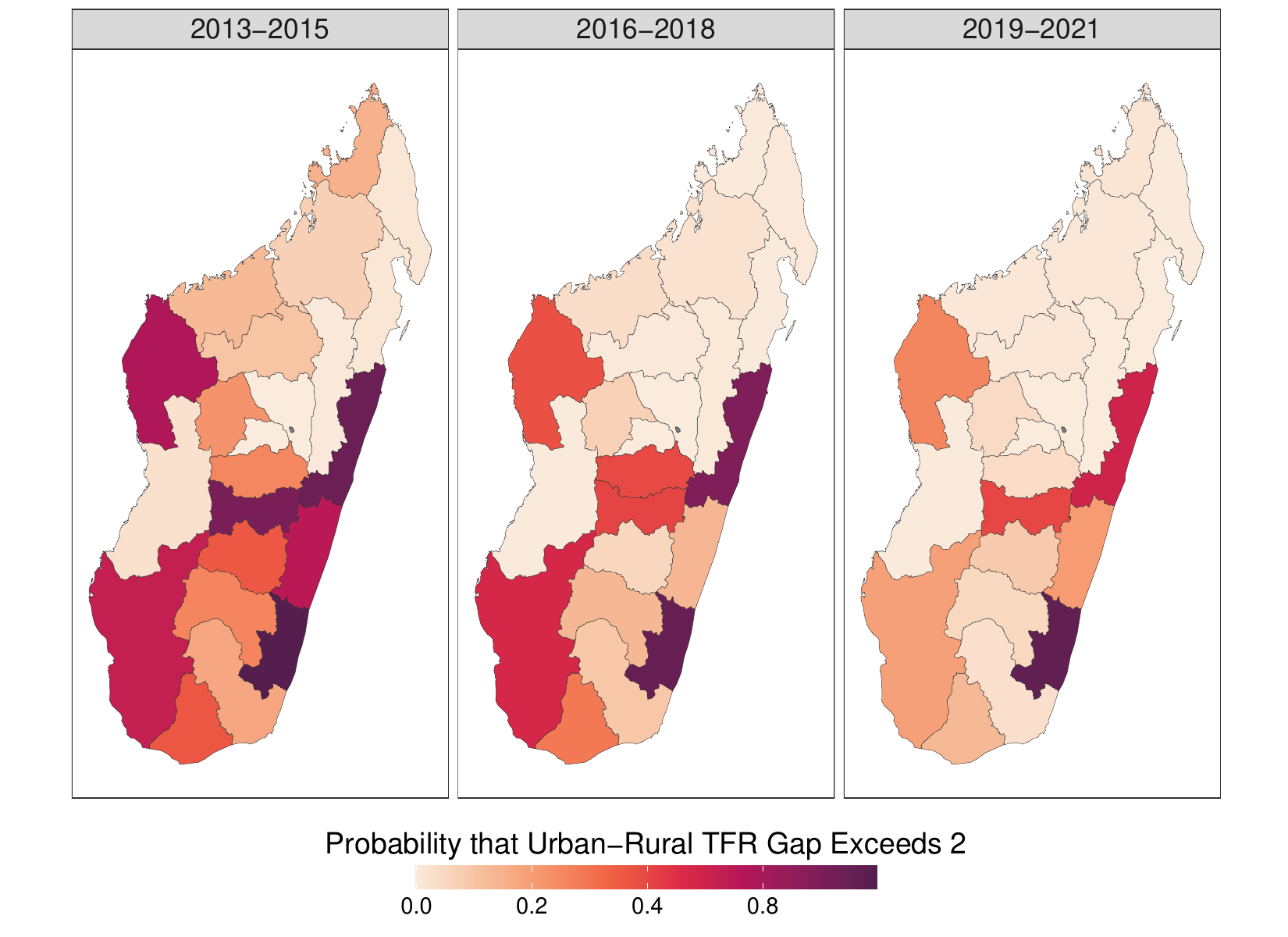}
    \caption{Probability that the urban-rural difference in TFR exceeds two children per woman, across Madagascar's Admin-1 regions, for three consecutive periods: 2013–2015, 2016–2018, and 2019–2021.}
    \label{fig:UR-diff-adm1-3yrs}
\end{figure}

Over time, fertility rates in both urban and rural populations are declining, but the rate of decline may vary considerably between the two, leading to persistent disparities. These differences reflect variations in demographic transitions, socioeconomic conditions, and access to essential services such as education, and family planning. To assess the extent of urban-rural fertility disparities, we examine the probability that the total fertility rate (TFR) in rural areas exceeds that of urban areas by more than two children per woman, estimated from the unit-level model with covariates.

Figure \ref{fig:UR-diff-adm1-3yrs} presents maps of this exceedance probability for Admin-1 regions in Madagascar across three consecutive 3-year periods up to 2021. While urban-rural fertility gaps have generally narrowed over time, as indicated by decreasing exceedance probabilities, substantial regional differences remain. In particular, the southeastern region of Atsimo Atsinanana continues to exhibit one of the highest probabilities (greater than 0.8) of exceeding the two-child difference threshold.

\subsection{Predictive Contribution of Covariates in Fertility Models}

We emphasize that the primary objective of this study is to predict fertility rather than to examine the underlying relationship between fertility and covariates. Maternal education and modern contraptive usage, are included as auxiliary predictors to improve model performance. This study does not seek to establish causal claims or draw inferential conclusions regarding the association between the covariates and fertility. 

 
We first assess the predictive contribution of the covariates within our unit-level model at the Admin-2 level. Separate models are estimated for urban and rural populations to account for differences in fertility patterns. In the urban model, areas with a 10\% higher attendance rate are predicted to have 9.7\% (95\% credible interval, 4.8\%, 14.2\%) lower fertility, on the population-average scale (with all other characteristics, i.e.,~covariates and random effects, being equal); regions with a 10\% higher modern contraceptive usage rate are predicted to have 7.5\% (-1.8\%, 15.5\%) lower fertility. In the rural model, the estimates were similar, such that areas with 10\% higher levels of education and contraceptive usage are also predicted to have lower fertility by 8.7\% (5.3\%, 11.9\%) and 9.7\% (5.9\%, 13.3\%), respectively.  


We further analyze the contributions of spatial, temporal, and maternal age-group effects, along with their interactions, to the total variance in the unit-level models at Admin-2 level. Additionally, we examine how the inclusion of covariates influences the magnitude of these effects by comparing two versions of the unit-level models across urban and rural settings. 

\begin{table}[ht]
\centering
\begin{tabular}{rrrrrr}
  \hline
Models & Time & Space & Age $\times$ Space & Space $\times$ Time & Age $\times$ Time \\ 
  \hline
Urban (cov) & 2.1\% & 83.7\% & 12.1\% & 1.0\% & 0.1\% \\ 
Urban (no cov) & 1.0\% & 91.5\% & 6.5\% & 0.6\% & 0.4\% \\ 
Rural (cov) & 10.3\% & 57.8\% & 25.2\% & 2.0\% & 4.7\% \\ 
Rural (no cov) & 4.0\% & 83.0\% & 10.4\% & 0.8\% & 1.8\% \\ 
   \hline
\end{tabular}
\caption{Variance decomposition across urban/rural unit-level models at Admin-2 level, differentiated by the inclusion of covariates. Age-group 45--49 is excluded, because the magnitude is very large and would dominate.}
\label{tab:var-decompose-model-UR-UYS}
\end{table}



The total variance in the model can be decomposed as the sum of the variance contributions from the random effects:
\begin{equation*}
\text{Var}(\eta_{i,t,a}) =\sigma^2_{\text{\tiny{S}}}+\sigma^2_{\text{\tiny{A}}}+\sigma^2_{\text{\tiny{T}}}+\sigma^2_{\text{\tiny{S,T}}}+\sigma^2_{\text{\tiny{S,A}}}+\sigma^2_{\text{\tiny{T,A}}}
\end{equation*}
where \( \sigma^2_{\text{\tiny{S}}} \), \( \sigma^2_{{\text{\tiny{A}}}} \), and \( \sigma^2_{\text{\tiny{T}}} \) 
represent the variance components associated with spatial (Admin-1) effects, maternal age group effects, and temporal effects, respectively. While \( \sigma^2_{\text{\tiny{S,T}}} \), \( \sigma^2_{\text{\tiny{S,A}}} \), and \( \sigma^2_{\text{\tiny{T,A}}} \) account for interactions between space-time, space-age, and time-age. These variance components are empirically estimated from the fitted random effects.

\begin{figure}[ht]
    \centering
    \includegraphics[clip, trim=0.1cm 0.2cm 0.2cm 0.1cm, width=1\linewidth]{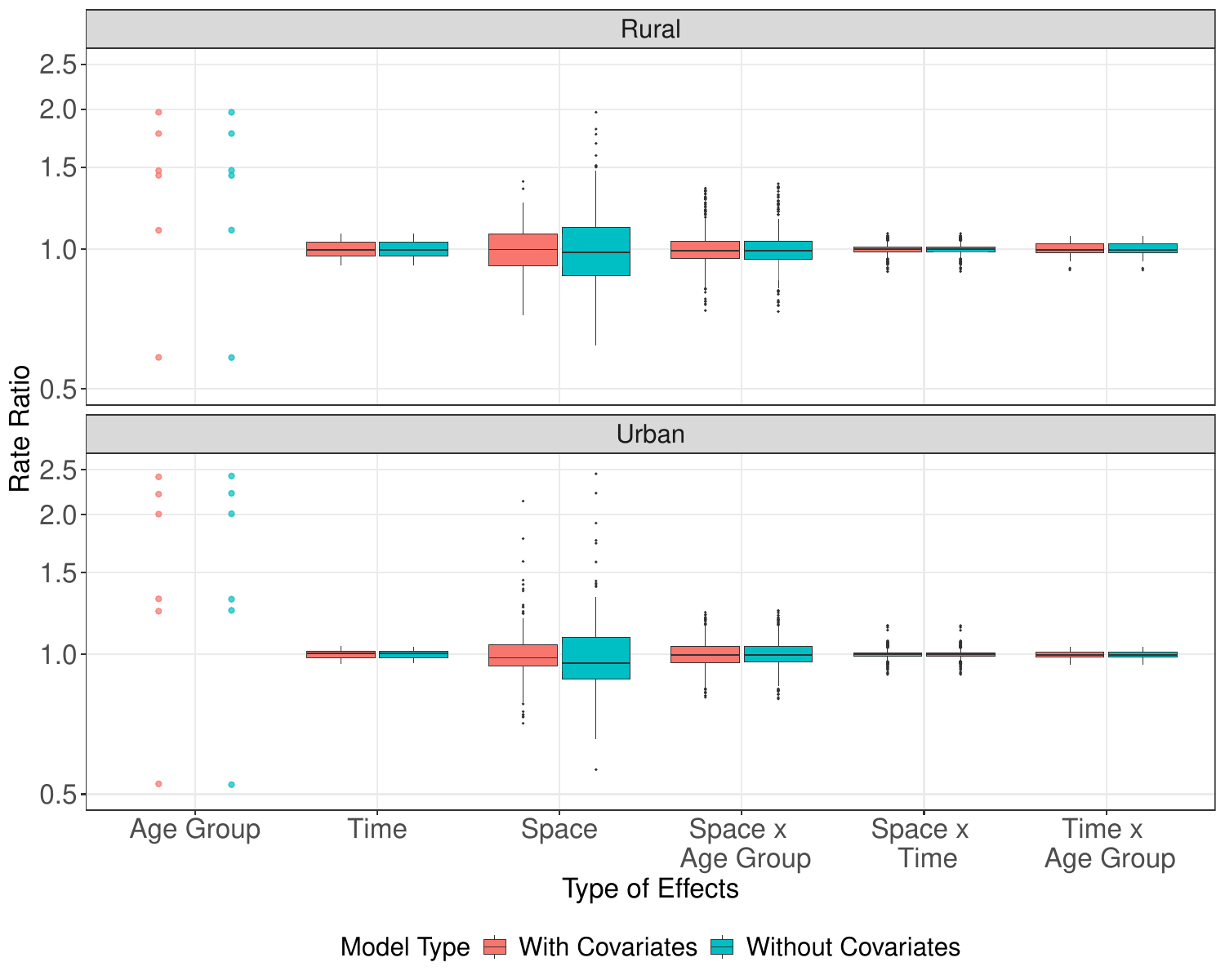}
\caption{Comparison of estimated random effects across urban/rural unit-level models at Admin-2 level, differentiated by inclusion of covariates.}
\label{fig:fert-ran-eff}
\end{figure}

Across all four models, the maternal age group main effects contribute the largest share of the variation. In the urban model with covariates, maternal age accounts for 97.4\% of the total variance, while in the rural model with covariates, it accounts for 96.3\%. In the urban model without covariates, maternal age explains 94.9\% of the variance, and in the rural model without covariates, it explains 91.1\%. 

The remaining variation is attributed to spatial, temporal, and interaction effects, with differences between urban and rural classifications and whether covariates are included, shown in Table \ref{tab:var-decompose-model-UR-UYS}. The remaining variation is primarily attributed to spatial random effects, which account for a smaller proportion in models that include education compared to their counterparts without education. This suggests that part of the spatial variation in fertility is captured by differences in educational attainment.

To further illustrate the impact of including education, Figure \ref{fig:fert-ran-eff} presents the distribution of estimated random effects across models. The random effects are plotted on an exponentiated scale so that the magnitudes reflect the fertility rate ratio relative to the overall mean, given the sum-to-zero constraint within each random effect group. The effects for the maternal age group 45–49 are excluded due to their disproportionately large magnitude, as fertility rates in this age group are significantly lower than other groups.

The patterns in Figure \ref{fig:fert-ran-eff} align with the variance decomposition in Table \ref{tab:var-decompose-model-UR-UYS}. While the inclusion of covariates has minimal impact on maternal age group and temporal effects, it substantially reduces spatial variability, as evidenced by the narrower distributions for these components in the models that include covariates.

\section{Model Validation}
\label{sec:fert-cross-validation}

\subsection{Cross-validation Setup}

We assess the validity of the area-level and unit-level models by testing their ability to recover ASFR and TFR estimates when data for specific combinations of maternal age group, Admin-1 area, and reference period are excluded. Model predictions for the leave-out combinations are then compared against direct estimates, which serve as the observed data summaries that we predict.

\begin{figure}[H]
  \centering
    \includegraphics[clip, trim=0cm 0cm 0cm 0cm, width=1\linewidth]{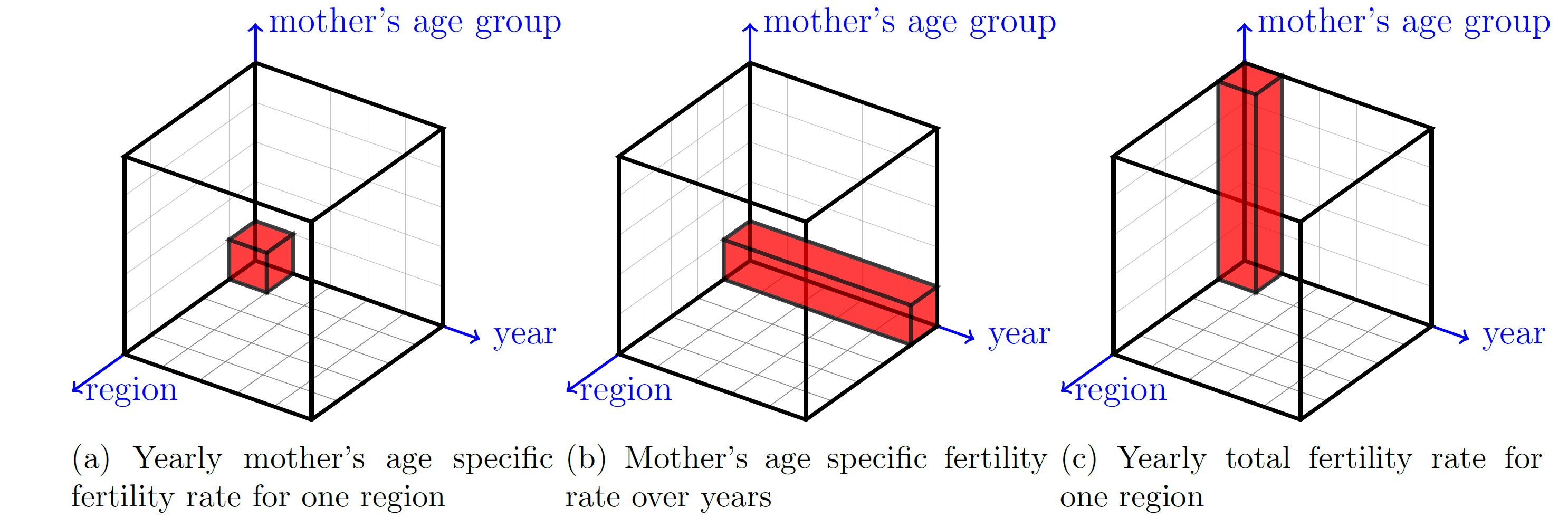}
\caption{The three dimensions for fertility rates} \label{fig:fert-three-dimenisons}  
\end{figure}

Our cross-validation scheme is designed to evaluate how well the models borrow information across related domains, particularly in capturing interactions between age, space, and time. Excluding entire regions would not adequately test this ability. On the other hand, due to data sparsity, direct estimates at the finest resolution—such as yearly ASFR estimates at the Admin-1 level, represented by the cube in Figure \ref{fig:fert-three-dimenisons}(a)—are highly unstable. This makes leaving out individual combinations at such a fine scale infeasible.

To balance these considerations, we adopt a leave-one-combination-out approach while leveraging data from the past 9 years to construct the models. We perform two rounds of cross-validation:

\begin{itemize}
\item \textbf{ASFR cross-validation:} We exclude data for a specific combination of Admin-1 area \(\times\) maternal age group across the full 9-year modeling period. Predicted ASFRs for the omitted combination are compared against the direct estimates obtained for the same Admin-1 \(\times\) age group, as represented by the rectangle in Figure \ref{fig:fert-three-dimenisons}(b).

\item \textbf{TFR cross-validation:} We exclude data for a specific combination of Admin-1 area \(\times\) 3-year period across all maternal age groups. Predicted TFRs are compared against the direct estimates for the same Admin-1 area \(\times\) 3-year period, as represented by the rectangle in Figure \ref{fig:fert-three-dimenisons}(c).
\end{itemize}

We validate models using data from the past 9 years. For both area-level  and unit-level model, we evaluate two versions:~one that includes area-level covariates as predictors and one without, for full details, see Section \ref{sec:covariate-model} in the supplemental materials. For the area-level model, we use the ASFR and TFR models outlined in Section~\ref{sec:fert-area-level-model} for ASFR and TFR validation, respectively.  For models incorporating temporal smoothing, we first obtain yearly estimates. In the case of TFR, these yearly estimates are then aggregated into 3-year periods, using population-based weights for each year.

Point estimates (posterior medians) of the log rates from the area-level or unit-level models based on the held-out data, are denoted 
\[
\widehat{\bm{\eta}} = (\widehat{\eta}_1, \dots ,\widehat{\eta}_B)^{\text{\tiny{T}}},
\] 
and the corresponding (comparison) direct estimates are
\[
\widehat{\bm{\eta}}^\text{W} = (\widehat{\eta}^\text{W}_1, \dots ,\widehat{\eta}^\text{W}_B),
\]
for Admin-1 areas \(\times\) maternal age group (or Admin-1 areas \(\times\) year) combinations, where \( B \) is the total number of area/age combinations.
Model performance is evaluated based on a collection of scoring rules and metrics, including relative bias, absolute bias, and root mean square error (RMSE), as defined below. 


\begin{eqnarray*}
\mbox{Bias}(\widehat{\bm{\eta}}, \widehat{\bm{\eta}}^\text{W}) &=& 100 \times\frac{1}{B} \sum_{b=1}^{B}  (\widehat{\eta}_b - \widehat{\eta}^\text{W}_b)  \\
\mbox{Absolute Bias}(\widehat{\bm{\eta}}, \widehat{\bm{\eta}}^\text{W}) &=& 100 \times \frac{1}{B} \sum_{b=1}^{B} 100 \times |\widehat{\eta}_b - \widehat{\eta}^\text{W}_b| \\
\mbox{Relative Bias}(\widehat{\bm{\eta}}, \widehat{\bm{\eta}}^\text{W}) &= & 100\% \times\frac{1}{B} \sum_{b=1}^{B}  \frac{\widehat{\eta}_b - \widehat{\eta}^\text{W}_b}{\widehat{\eta}^\text{W}_b} \\
\mbox{Absolute Relative Bias}(\widehat{\bm{\eta}}, \widehat{\bm{\eta}}^\text{W}) &= &100\% \times \frac{1}{B} \sum_{b=1}^{B}  \left|\frac{\widehat{\eta}_b - \widehat{\eta}^\text{W}_b}{\widehat{\eta}^\text{W}_b}\right| \\
\mbox{RMSE}(\widehat{\bm{\eta}}, \widehat{\bm{\eta}}^\text{W}) &=& \sqrt{\frac{1}{B} \sum_{b=1}^{B} (\widehat{\eta}_b - \widehat{\eta}^\text{W}_b)^2}
\end{eqnarray*}


In addition to comparisons of point estimate, we assess interval estimates using the frequentist coverage and interval score (IS). For example, the log-transformed direct estimates for a given combination \( i,t,a \) follows the sampling distribution,
\[
\widehat{\eta}^{\text{W}}_{i,t,a} \mid \eta_{i,t,a} \sim N(\eta_{i,t,a}, {V}_{i,t,a}),
\]
where the design-based variance estimate \( V_{i,t,a} \) is obtained via the Jackknife repeated replication method combined with the delta method, as described in Section~\ref{sec:fert-area-level-model}. 

To evaluate the frequentist coverage of interval estimates, we construct confidence intervals \( (l_b, u_b) \) using the \( \alpha/2 \) and \( 1 - \alpha/2 \) quantiles of the posterior predictive distribution of \( \widehat{\eta}_b \), where \( b \) indexes each area, time, and age group combination $(i, t, a)$. To account for uncertainty from the direct estimates, we add an error term \( \epsilon_b \sim N(0, \widehat{V}^{\text{W}}_b) \). The resulting interval  \( (l_b, u_b) \) is derived from these empirical quantiles, and coverage is evaluated by checking whether the direct estimate \( \widehat{\eta}^{\text{W}}_{b} \) falls within the constructed interval.

The interval score, as defined by \cite{gneiting2007strictly}, is computed as:  
\begin{eqnarray*}
\mbox{IS}_\alpha(\bm{l},\bm{u})&=& \frac{1}{B} \sum_{b} \left[u_{b} - l_{b} + \frac{\alpha}{2} (l_{b} - \widehat{\eta}^{\text{W}}_{b}) I(l_{b} > \widehat{\eta}^{\text{W}}_{b})\right. - \left.\frac{\alpha}{2} (u_{b} - \widehat{\eta}^{\text{W}}_{b})  I(u_{b} < \widehat{\eta}^{\text{W}}_{b}) \right].
\end{eqnarray*}
This metric rewards narrow intervals while penalizing intervals that fail to cover the estimates for the held-out observations. Lower scores indicate better model performance.


\begin{table}[ht]
\centering
\begin{tabular}{lrrrrrrr}
  \hline
  Method & Bias & \makecell[l]{Absolute \\ Bias} & \makecell[l]{Relative \\ Bias} & 
  \makecell[l]{Absolute \\ Relative \\ Bias} & RMSE & \makecell[l]{Coverage \\ (90\% CIs)} \\ 
  \hline
  \multicolumn{7}{l}{\textit{\textbf{Validate 3-year TFR}}} \\
  Unit (with cov) & 0.01 & 0.08 & 1.31\% & 5.48\% & 0.10 & 88\% \\ 
  Unit (no cov) & 0.01 & 0.08 & 1.37\% & 5.48\% & 0.10 & 88\% \\ 
  Area-level (with cov) & 0.00 & 0.07 & 0.32\% & 4.86\% & 0.09 & 90\% \\ 
  Area-level (no cov) & 0.00 & 0.07 & 0.36\% & 4.89\% & 0.09 & 88\% \\ 
  \multicolumn{7}{l}{\textit{\textbf{Validate 9-year ASFR}}} \\
  Unit (with cov) & 0.02 & 0.11 & -0.25\% & 5.73\% & 0.15 & 88\% \\ 
  Unit (no cov) & 0.02 & 0.11 & -0.23\% & 5.67\% & 0.15 & 90\% \\ 
  Area-level (with cov) & 0.02 & 0.15 & -0.37\% & 7.72\% & 0.20 & 84\% \\ 
  Area-level (no cov) & 0.02 & 0.15 & -0.39\% & 7.56\% & 0.20 & 86\% \\ 
  \hline
\end{tabular}
\caption{Summary of cross-validation results for different fertility models. ``With cov'' refers to models including covariates (contraceptive use and education indicators), while ``No cov'' refers to models without covariates.}
\label{tab:fertility_cv_summary}
\end{table}

\subsection{Results}

We present cross-validation results for fertility estimates based on the 2021 Madagascar DHS, using the models and metrics described above. Recall that Madagascar consists of 23 Admin-1 regions, which leads to 69 validation combinations for TFR across three 3-year reference periods. For ASFR validation, we exclude the 45-49 age group due to zero values in the direct estimates, resulting in 138 validation combinations. All results are reported on the log scale to provide a more stable measure for model performance, as rate estimates at the original scale can be small, with skewed sampling distributions. 

\begin{figure}[!ht]
    \centering
    \includegraphics[clip, trim=0.2cm 1.5cm 0.2cm 1cm, width=1\linewidth]{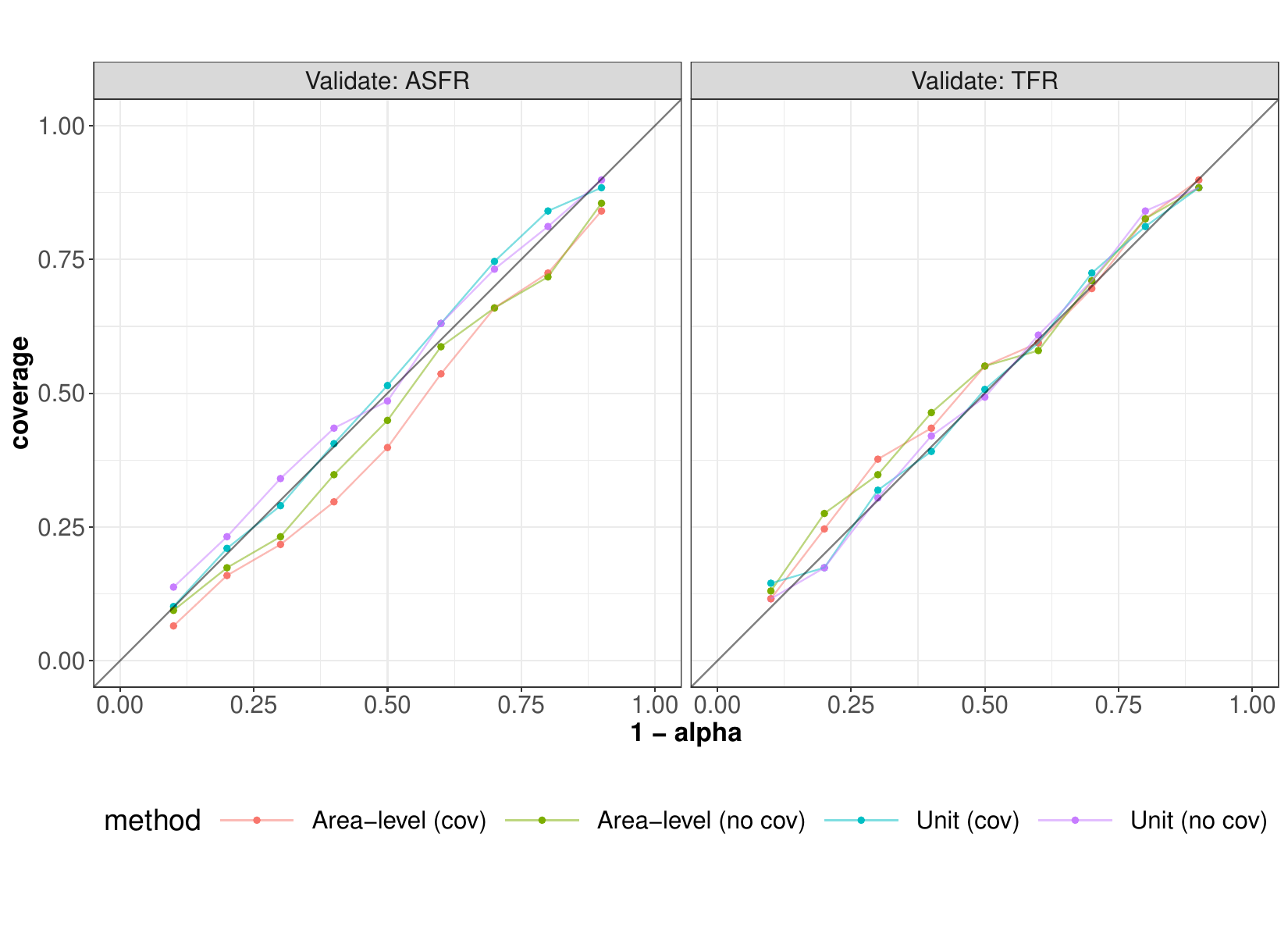}
    \caption{Coverage probability across different confidence levels for ASFR and TFR validation, based on the four methods.}
    \label{fig:fert-coverage-compare}
\end{figure}

Table \ref{tab:fertility_cv_summary} summarizes the cross-validation results for ASFR and TFR across different models. Bias estimates are close to zero across all models, indicating no systematic over- or under-estimation. For TFR validation, the area-level model achieves the lowest RMSE (0.09) and absolute bias (0.07), slightly outperforming both versions of the unit-level models. For ASFR validation, the unit-level models outperform the area-level model, with RMSE values of 0.15 compared to 0.20 for the area-level model. These results indicate that unit-level models provide better flexibility in capturing finer-scale fertility patterns on mother's age group. The inclusion of covariates does not significantly affect accuracy for point estimates, as both pairs of models yield similar results across all evaluation metrics.

Beyond point estimates, we assess interval estimates using coverage probability and interval scores. Figure \ref{fig:fert-coverage-compare} presents coverage probabilities across different  levels. In TFR validation, all models achieve nominal coverage. However, in ASFR validation, the area-level model systematically exhibits under coverage  reaching, for example, only 84\% coverage at the 90\% level for the version included covariates, whereas unit-level models maintain better coverage. This indicates that the area-level model might underestimate uncertainty in ASFR estimates.

\begin{figure}[!ht]
    \centering
    \includegraphics[clip, trim=0.2cm 1cm 0.2cm 0cm, width=1\linewidth]{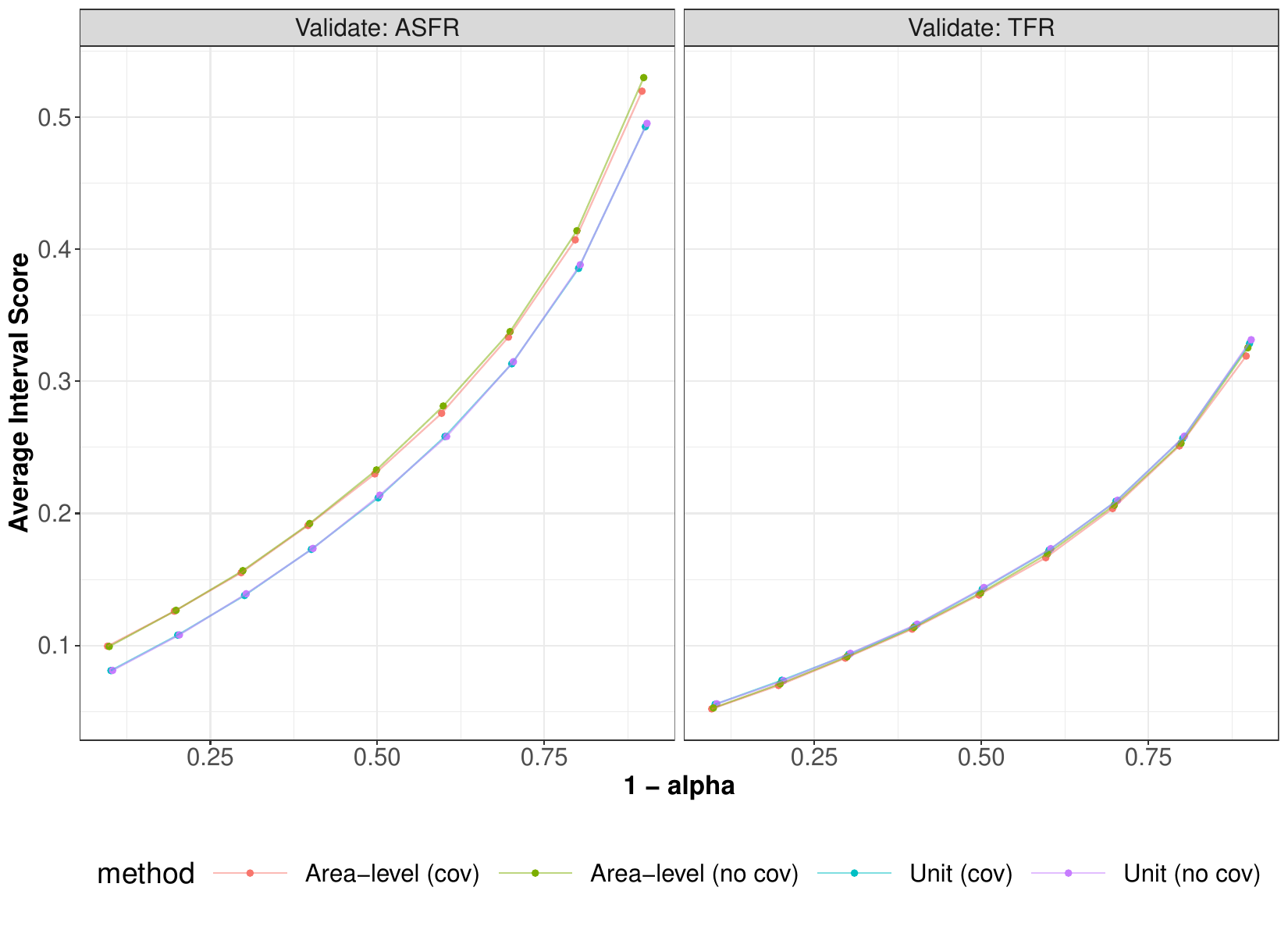}
    \caption{Interval score comparisons across different confidence levels for ASFR and TFR validation. Lower scores indicate more precise and well-calibrated uncertainty estimates.}
    \label{fig:fert-interval-score-compare}
\end{figure}

Figure \ref{fig:fert-interval-score-compare} shows interval scores, which measure the trade-off between interval width and accuracy. Models with lower values are preferred. In TFR validation, all models achieve similar interval scores; whereas in ASFR validation, unit-level models perform better, with lower interval scores, confirming the under-coverage we found for area-level models.

Overall, all four models perform reasonably well. The slightly worse performance of the area-level model in ASFR validation suggests that greater modeling flexibility, as provided by the unit-level models, is beneficial for capturing finer-scale fertility patterns. The comparison between the two versions of the unit-level model shows no significant difference in predictive accuracy, indicating that incorporating covariates does not substantially improve performance in this setting. One possible explanation is area-level covariates primarily captures spatial variation, but since our cross-validation scheme does not fully exclude data from entire regions, its predictive contribution is limited. 

\section{Discussion} 
\label{sec:fert-discussion}

In this study, we developed a comprehensive SAE framework for fertility estimation using DHS data, addressing data sparsity across space, time, and maternal age groups. Estimating ASFR and TFR at fine resolutions is challenging due to small sample sizes. Our framework provides a statistically rigorous workflow that integrates direct estimation, area-level models, and unit-level models, offering flexible solutions for different estimation needs. Direct and area-level models maintain design consistency and perform well with sufficient data, such as for Admin-1 TFR, while unit-level models borrow strength across space, time, and maternal age groups to produce more stable estimates at finer resolutions, such as yearly Admin-2 ASFR. This suite of models supports robust fertility estimation across geographic, temporal, and demographic (maternal age groups) scales.

To assess the reliability of our estimates, we conducted cross-validation, which confirms that our models provide reliable uncertainty estimates with reasonable interval coverage. Moreover, we demonstrate the practical utility of our framework through an application to the 2021 Madagascar DHS, illustrating how our models can be used to produce fertility estimates at subnational levels. 

Our approach also enhances fertility estimation by incorporating socioeconomic predictors, such as maternal education (measured as high school attendance rate), prevalence of modern contraceptive use and urban-rural location. Our models leverage these covariates to improve predictive accuracy at fine scales. Additionally, our model enables the tracking of urban-rural fertility disparities across regions and time, capturing socioeconomic influences on fertility patterns. By integrating these factors into the SAE framework, we provide a more nuanced understanding of fertility disparities, offering the possibility to enable targeted policy interventions to address regional and demographic variations in fertility trends.


A limitation arises from the independence assumption in our modeling framework. Our models treat births as a Poisson outcome with mother-year exposure as an offset, inherently assuming independence across mother-year periods. However, this assumption is violated because we track the same set of mothers retrospectively based on their birth history, rather than independently sampling mother-years, introducing a positive correlation structure that is not explicitly modeled. Additionally, births are a recurrent process with postpartum periods of no births, leading to negative correlation over adjacent periods from the same mother. 

There are potential improvement on the models we proposed to account for the correlation of outcomes on the same mother. For direct estimation, marginal variance remains valid since we use jackknife resampling, which also allows for empirical estimation of covariance among direct estimates. In area-level models, this covariance can be incorporated by modeling direct estimates as a multivariate normal distribution, rather than assuming independence. Unit-level models pose a greater challenge, as correlation cannot be directly estimated via resampling and must instead be specified through an appropriate data-generating process. While failing to model these dependencies affects uncertainty quantification, our cross-validation results in Section \ref{sec:fert-cross-validation} suggest that model coverage remains reasonable, indicating that the impact on uncertainty estimation, if present, may not be significant. We do note that  issue stemming from using multiple outcomes on the same mothers are common to the estimation of under-5 mortality (which is also based on retrospective birt histories), and these aspects are routinely ignored in that literature, see for example, \cite{li:etal:19}.

\bibliographystyle{natbib} 
\bibliography{refs}

\newpage 
\setcounter{section}{0}
\setcounter{figure}{0}
\setcounter{equation}{0}

\makeatletter
\renewcommand \thesection{S\@arabic\c@section}
\renewcommand \thefigure{S\@arabic\c@figure}
\renewcommand \theequation{S\@arabic\c@equation}

\makeatother

\centerline{{\huge Supplemental Materials} }

\section{Modeling Covariates}
\label{sec:covariate-model}

The covariates included in the model are two indicators, both defined using women aged 15-49 as the denominator, which aligns with the target population for fertility estimation. The first is the proportion of women who have ever attended lower secondary school or higher. The second is the prevalence of modern contraceptive use, defined to include methods such as sterilization, pills and condoms etc., and other recognized modern methods as classified in the DHS. These indicators were selected due to their strong relevance to fertility, as discussed in the main manuscript.

We aim to provide area-level estimates for these two covariates to be used in the fertility model, both at Admin-1 and Admin-2 models. We adopt an area-level Fay-Herriot (FH) model \citep{fay:herriot:79} to estimate these indicators. Let $\widehat{p}_i^{\,\text{W}}$ denote the weighted direct estimate of high school attendance rate or contraceptive prevalence for area $i$, and $V_i$ its corresponding variance. Since the outcome is binary and constrained to $(0,1)$, we apply a logit transformation 
$$
\text{logit}\left(\widehat{p}_i^{\,\text{W}}\right) \mid p_i \sim N(p_i, V_i),
$$
where $p_i$ represents the true underlying logit-transformed prevalence for area $i$. $\theta_i$s are then modeled through a hierarchal mean model with a linear mixed-effects structure:
$$
\theta_i = \alpha + u_i,
$$
where $\alpha$ is an intercept and $u_i$ is a spatial random effect.


To account for both structured and unstructured spatial effect, we adopt the BYM2 formulation \citep{riebler2016an}:
\[
u_i = \sigma \left( \sqrt{1 - \phi} \, e_i + \sqrt{\phi} \, S_i \right),
\]
where $e_i \overset{\text{iid}}{\sim} N(0,1)$ and $S_i$ follows a scaled intrinsic conditional autoregressive (ICAR) prior \citep{besag1991bayesian}. Penalized complexity (PC) priors \citep{simpson2017penalising} are placed on the total variation $\sigma$ and proportion contributed by the structured effect $\phi$.

\newpage
\section{Urban/Rural Stratification}

\subsection{Urban-Rural Stratification in Fertility Modeling}

When modeling fertility outcomes, it is important to adjust for urban/rural (U/R) stratification if the outcome of interest differs by urban/rural status and the survey design involves unequal sampling across urban and rural clusters \citep{wu2024modelling}. 

To properly account for U/R stratification, we follow a pipeline, summarized in Figure~\ref{fig:UR-pipeline-overview}. We first obtain separate fertility estimates by fitting U/R specific models. Overall estimates for fertility are then derived as weighted combinations of these U/R specific estimates, using the urban fraction $r_i$ as the weight for area $i$. In the Bayesian setting, this aggregation is applied to the posterior samples, such that naturally propagate uncertainty from the U/R specific models into the combined estimates.

\tikzstyle{decision} = [diamond, draw, fill=blue!20, 
    text width=5.5em, text badly centered, node distance=3cm, inner sep=0pt]
\tikzstyle{block} = [rectangle, draw, fill=blue!20, 
    text width=10em, text centered, rounded corners, minimum height=4em]
\tikzstyle{line} = [draw, -latex']
\tikzstyle{cloud} = [draw, ellipse,fill=red!20, node distance=3cm, text width=4.5em,text centered,
    minimum height=2em]
    
\begin{figure} [!ht]

     \begin{center}
    \begin{small}
    
\begin{tikzpicture}[node distance = -.3cm, auto]
    \node [block] (level1left) {Education\\ Model};
        \node [block, xshift=5.5cm] (level1right) {U/R Classification, \\ Fraction Model};
    \node [block, below of=level1left, yshift=-3cm, xshift=2.75cm] (level2) {Aggregation\\ Model};

    \path [line] (level1left) -- (level2);
        \path [line] (level1right) -- (level2);
   
\end{tikzpicture}

    \caption{Overview of modeling process}

\label{fig:UR-pipeline-overview}

\end{small}
     \end{center}
\end{figure}
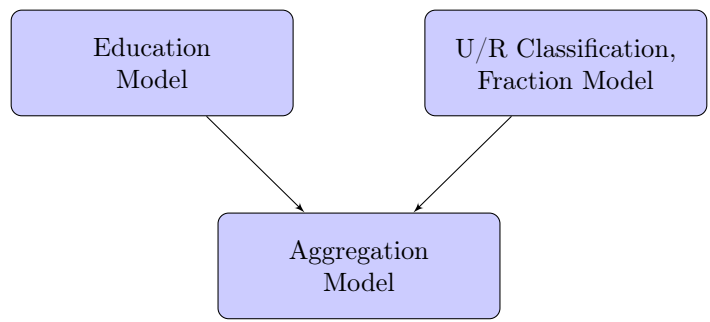

The key input for this aggregation step is the urban fraction $r_i$, representing the proportion of the female population (aged 15--19, 20--24, etc.) residing in urban areas according to the survey's sampling frame. Estimating $r_i$ directly at fine spatial resolutions (e.g., Admin-2) is often infeasible due to limited sample sizes. To overcome this, we adopt the modeling pipeline proposed in \citet{wu2024modelling}, with minor simplifications.

Specifically, we pixelate the study region and assign each pixel an urban or rural label based on a trained classification model. Population density data from WorldPop \citep{stevens2015disaggregating} are used to distribute the relevant female population across these pixels. This allows for the estimation of urban fractions in any arbitrary region by combining the classification and population rasters.

Formally, let $L$ denote a binary raster map indicating urban status, with $L_g = 1$ if pixel $g$ is classified as urban and $L_g = 0$ otherwise. Let $H$ represent the population density map over the same grid, and $H_g$ denote the estimated number of women population for specific age group(s) in pixel $g$. Then, the urban fraction in a given region is computed as:
\begin{equation}
\label{eq:supp-ur-frac}
    r = \frac{\sum_g L_g \, H_g}{\sum_g H_g}.
\end{equation}

\newpage

\section{Additional Visualizations}
\subsection{Admin-2 Urban-Rural Disparities in TFR}

\begin{figure}[ht]
    \centering
    \includegraphics[clip, trim=0.1cm 0.5cm 0.2cm 0.2cm, width=1\linewidth]{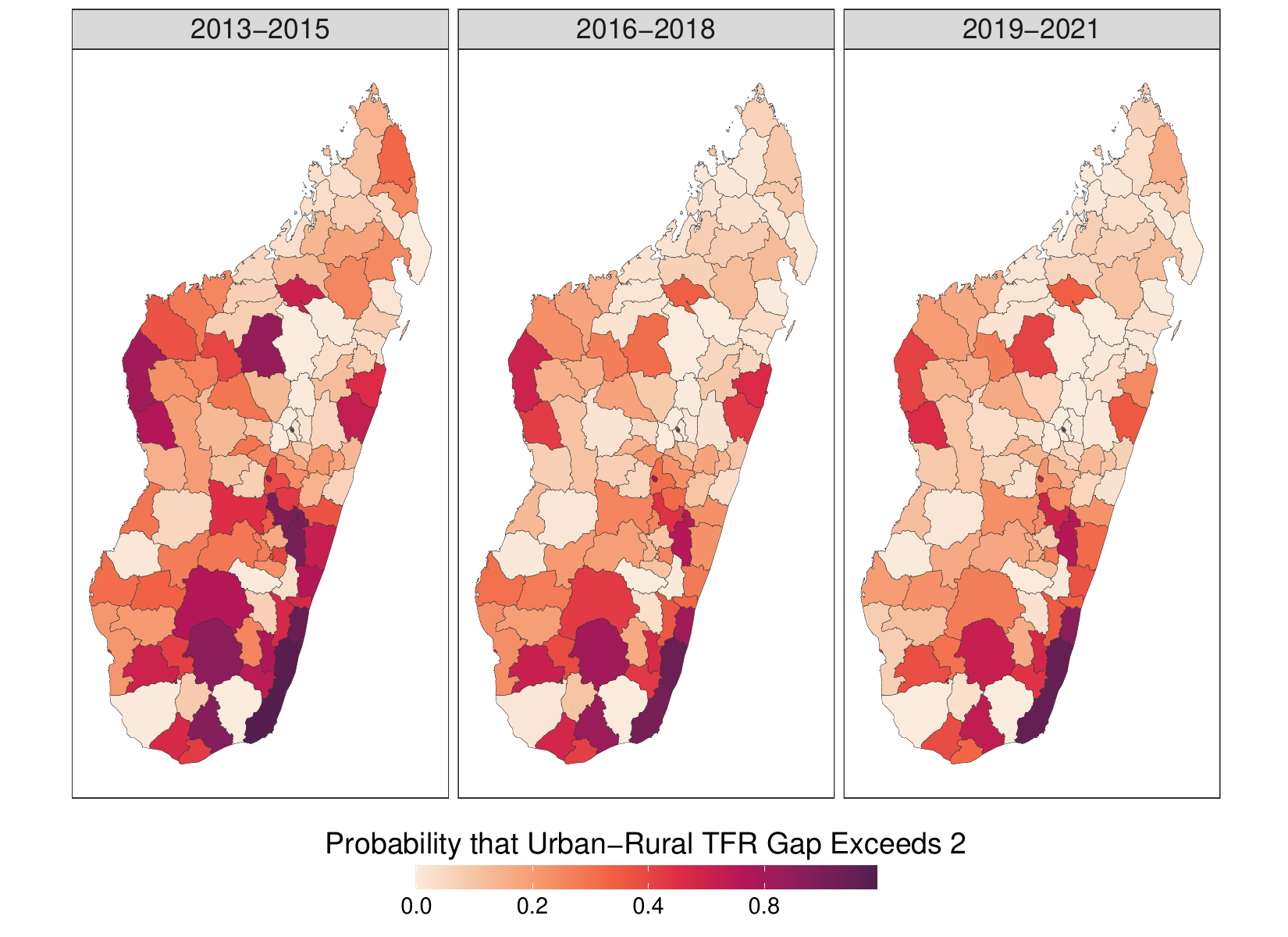}
    \caption{Probability that the urban-rural difference in TFR exceeds two children per woman, across Madagascar's Admin-2 regions, for three consecutive periods: 2013--2015, 2016--2018, and 2019--2021. Results are based on Admin-2 level unit-level models with covariates.}
    \label{fig:UR-diff-adm2-3yrs}
\end{figure}

\newpage
\subsection{Interval Plots for Admin-1 TFR}

\begin{figure}[ht]
    \centering
    \includegraphics[clip, trim=0.1cm 0.2cm 0.2cm 0.2cm, width=1\linewidth]{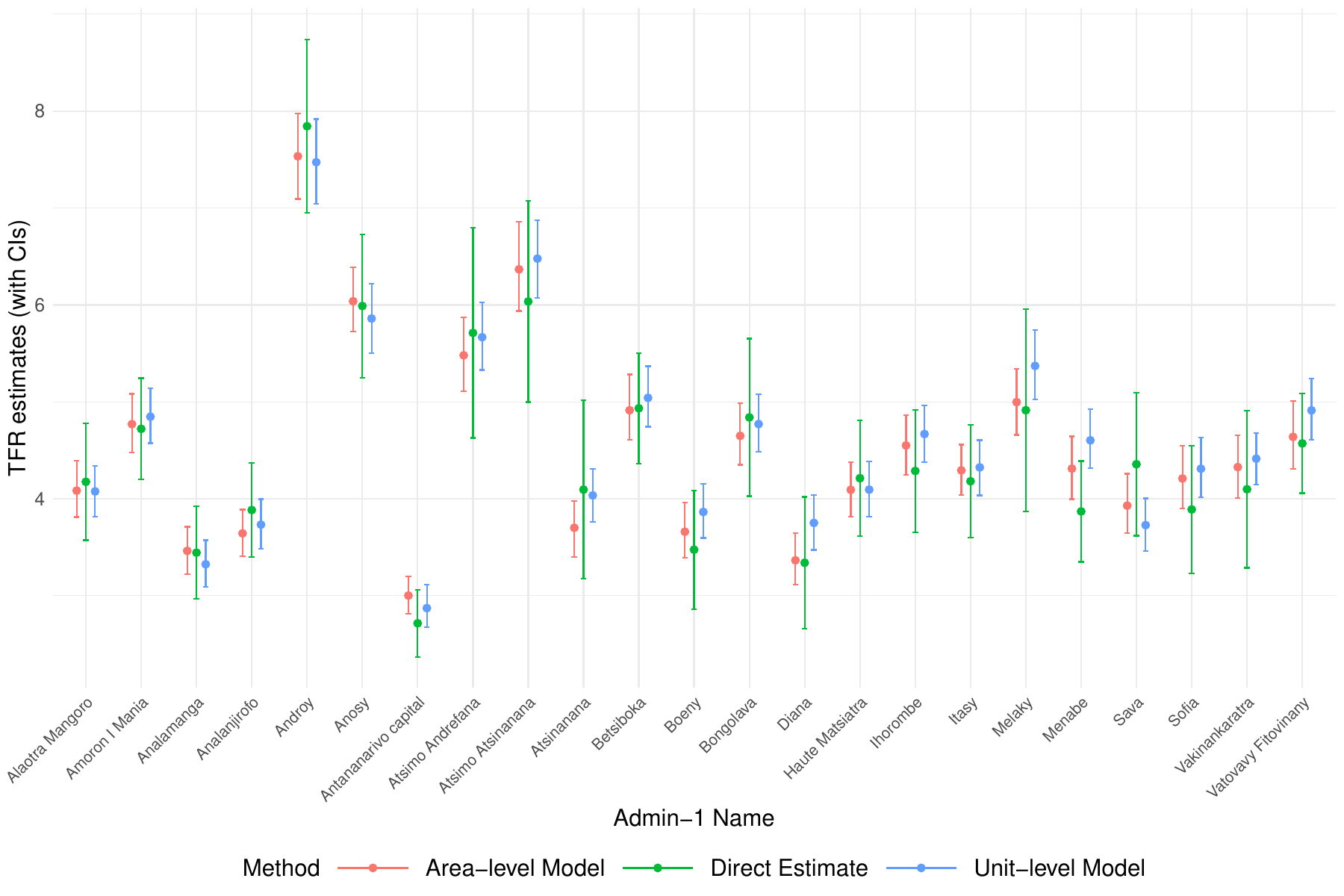}
    \caption{Admin-1 estimates (with 90\%CIs) for TFR in Madagascar for 2019--2021. Area-level and unit-level models are the versions with covariates.}
    \label{fig:TFR-interval-adm1}
\end{figure}

\end{document}